\documentclass[3p,number,sort&compress,times,fleqn]{elsarticle}   

\usepackage[colorlinks = true,
linkcolor = blue,
urlcolor = blue,
citecolor = blue,
anchorcolor = blue]{hyperref}

\usepackage{graphicx}
\usepackage[labelfont=bf, justification=justified]{caption}
\usepackage{algorithm}
\usepackage{algpseudocode}

 \usepackage[table,xcdraw]{xcolor}
 \usepackage[braket, qm]{qcircuit}

\usepackage{style/csml}
\usepackage{braket}

\graphicspath{{./figs/}}
\usepackage{url}
\usepackage{setspace}
\DeclareMathOperator\erf{erf}

\begin{document}
	
\begin{frontmatter}
	
\title{Towards Quantum Computational Mechanics}

\author[1]{Burigede Liu}
\ead{bl377@cam.ac.uk}
\author[2,3]{Michael Ortiz}
\ead{ortiz@aero.caltech.edu}
\author[1]{Fehmi Cirak}
\ead{fc286@cam.ac.uk}


\address[1]{Department of Engineering, University of Cambridge, Trumpington Street, Cambridge, CB2 1PZ, UK }
\address[2]{Division of Engineering and Applied Science, California
Institute of Technology, Pasadena, CA 91125, USA}
\address[3]{Institut f\"ur Angewandte Mathematik and 
Hausdorff Center for Mathematics, 
Universit\"at Bonn, Endenicher Allee 60, 53115 Bonn, Germany}

\begin{abstract}
The advent of quantum computers, operating on entirely different physical principles and abstractions from those of classical digital computers, sets forth a new computing paradigm that can potentially result in game-changing efficiencies and computational performance. Specifically, the ability to simultaneously evolve the state of an entire quantum system leads to quantum parallelism and interference. Despite these prospects, opportunities to bring quantum computing to bear on problems of computational mechanics remain largely unexplored. In this work, we demonstrate how quantum computing can indeed be used to solve representative volume element (RVE) problems in computational homogenisation with polylogarithmic complexity of~$ \mathcal{O}((\log N)^c)$, compared to~$\mathcal{O}(N^c)$ in classical computing. Thus, our quantum RVE solver attains exponential acceleration with respect to classical solvers, bringing concurrent multiscale computing closer to practicality. The proposed quantum RVE solver combines conventional algorithms such as a fixed-point iteration for a homogeneous reference material and the Fast Fourier Transform (FFT). However, the quantum computing reformulation of these algorithms requires a fundamental paradigm shift and a complete rethinking and overhaul of the classical implementation. We employ or develop several techniques, including the Quantum Fourier Transform (QFT), quantum encoding of polynomials, classical piecewise Chebyshev approximation of functions and an auxiliary algorithm for implementing the fixed-point iteration and show that, indeed, an efficient implementation of RVE solvers on quantum computers is possible. We additionally provide theoretical proofs and numerical evidence confirming the anticipated~$ \mathcal{O} \left ((\log N)^c \right) $ complexity of the proposed solver. 
\end{abstract}
	

\end{frontmatter}


%
\section{Introduction \label{sec:intro}}
%

\subsection{Why quantum computing?}
The phenomenal advances in computational mechanics from its early beginnings in the late 1950s to today follow an exponential increase in computing power over the same period of time~\cite{ shalf2020future,national2019quantum}. This progress is most evident from the increase of finite element problem sizes from less than hundreds ($<10^2$) to hundreds of billions~($>10^{11}$) of elements over the same time period~\cite{clough1990original, kohl2023fundamental}. As is widely known, Moore's law stipulates a two-fold increase in the density of transistors on a circuit every two years. However, the consensus is that future improvements in computing power will be limited because of the physical limitations in shrinking logic gate sizes further. A naive scaling-up of current computing technologies without an increase in transistor densities is also believed to be highly problematic because of excessive cost and energy consumption. Transitioning to quantum systems to perform computations brings about a radically different computing paradigm and a possible avenue out of this impasse. 

\subsection{The origins of quantum computing.}
Quantum computers were envisaged in early 1980s by Feynman~\cite{feynman1982simulating}, Benioff~\cite{benioff1980computer} and Manin~\cite{manin1980computable}. Feynman was motivated by the impossibility of simulating a reasonably-sized \emph{quantum system} with classical computers. For instance, the state of a quantum system consisting of~$n_p$ particles contained in a box in~$\mathbb R^3$ is described by a complex-valued \emph{wave function}~$\psi(\vec x) : \mathbb R^{3 n_p} \mapsto \mathbb C $. The discretisation of this configuration space with only ten cells in each dimension leads to a discretised system with~$10^{3n_p}$ degrees of freedom, rendering the problem unsolvable beyond a few particles. Feynman argued that the efficient simulation of quantum systems requires non-classical computers that exploit phenomena unique to quantum mechanics. The same argument was put forward by Manin in his book in Russian; see~\cite{preskill2023quantum}. 

In contrast to both, Benioff was concerned with energy dissipation in classical computers resulting from the irreversibility of elementary operations. For instance, in a classical computer an~$AND$ gate has two binary inputs and one binary output and is irreversible, as it is impossible to deduce from the single output the corresponding inputs. This irreversibility in turn necessarily entails a loss of energy~\cite{landauer1961irreversibility, bennett1973logical}. Benioff proposed an alternative quantum-based classical computer design in which every operation is encoded as the solution of the reversible time-dependent Schr\" odinger equation of quantum mechanics. 

The notion of a quantum computer, as envisaged by Feynman and Manin, was finally formalised in the seminal work by Deutsch~\cite{deutsch1985quantum} a few years later. In the same work, Deutsch also discussed the advantages of a quantum computer in solving problems beyond the simulation of quantum systems, which led to an initial flurry of quantum algorithms, including Grover's algorithm of unstructured search~\cite{grover1996fast} and Shor's algorithm for prime factorisation~\cite{ shor1999polynomial}.

\subsection{A different way of thinking.}
To develop an intuitive understanding of quantum computing, adopting a physical viewpoint of computing is expedient. Briefly, computing is always tied to a physical representation and a computer is a machine. As Deutsch~\cite{deutsch1985quantum} stated: \emph{"a computing machine is any physical system whose dynamical evolution takes it from one of a set of 'input' states to one of a set of 'output' states".} The input state of a quantum computer is chosen from the configuration space of a quantum system. According to the postulates of quantum mechanics, the configuration of the system with~$n$ quantum particles is described by a \emph{state vector}~$\ket{q}$, see~\cite{sakurai1995modern}.\footnote{The state vector~$\ket q$ and wave function~$\psi(\vec x)$ are equivalent representations of a quantum system. In quantum computing usually only state vectors are used.} The ket symbol~$\ket{\cdot}$ merely indicates that q is a quantum mechanical vector. Each particle is referred to as a \emph{qubit} and has two states. These two states can be, for instance, the up and down spin states of a spin 1/2 particle, ground and excited states of an atom, or the horizontal and vertical polarisation of a photon~\cite{divincenzo2000physical}. Owing to \emph{quantum entanglement}, the state vector of~$n$ qubits is the Hilbert space~$\mathbb C^{2^n}$, i.e.~$\ket q \in \mathbb C^{2^n}$. 
Note that this space is exponentially larger than the configuration space of~$n$ classical particles which is only~$\mathbb C^{2n}$, see~\cite{jozsa2003role}. The time evolution of the state vector~$\ket {q (t)}$ is governed by the Schr\"odinger equation and the state vector~$\ket {q(T)} $ at time~$t=T$ is given by a linear mapping~$U_C: \ket{q(0)} \mapsto \ket{q(T)}$. 

The evolution operator~$U_C \in \mathbb C^{2^n \times 2^n}$, or {\sl propagator}, is {\sl unitary}, $U_C^{-1} = U_C^\dagger$, i.~e., it is length preserving and bijective. A \emph{universal quantum computer} can  implement any such unitary operator~$U_C$, i.~e., it performs computation by evolving a chosen~$\ket {q(0)}$ to~$\ket {q(T)}$ using an arbitrary, user-designed unitary~$U_C$. Because of these characteristics, quantum computers are sometimes regarded as \emph{analog machines}, and their ability to simultaneously evolve the state vector~$\ket {q(t)}$ in one fell swoop as \emph{quantum parallelism}. 

In applications to mechanics, the abstract description of quantum computing provided so far requires further elaboration. For instance, in computational mechanics the initial state~$\ket {q(0)}$ may represent the forcing, the final state~$\ket {q (T)}$ the solution and~$U_C$ the discretised inverse solution operator. A classical forcing vector~$\vec q \in \mathbb C^N$, with~\mbox{$N=2^{n}$}, is encoded as the components, called~\emph{amplitudes},~$q_k (0) \in \mathbb C$ of a state vector~$\ket{q(0)} = \sum_k q_k(0) \ket k$ of a system with~$n$ qubits. The basis vectors~\mbox{$\ket k \in \{ \ket 0 , \, \ket 1, \dotsc \ket{2^n-1} \}$} provide a labelling of the~$2^n$ possible states of the quantum system. This encoding is referred to as \emph{state preparation} and can be represented by the unitary mapping~\mbox{$U_I(\vec q) \colon \ket { 0 \dotsc 0} \mapsto \ket {q(0)}$}. Here,~$\ket { 0 \dotsc 0} $ is the known initial state and~$U_I(\vec q)$ a unitary matrix depending on the classical forcing data. Algorithms for state preparation can be found, e.g., in~\cite{mottonen2004transformation, shende2005synthesis, lubasch2020variational,araujo2021divide}. 

A subsequent application of the propagator $U_C$, representing the mechanical system under consideration, performs the desired computation. 
It is well-known that any unitary matrix can be represented as the composition of a few elementary unitary matrices~\cite{deutsch1989quantum, barenco1995elementary}. These elementary unitaries are of dimension~$2\times2$ and~$4\times4$ and are applied to a single qubit or to two qubits at a time, respectively. In gate-based quantum computing the composition of the unitaries~$U_I$ and~$U_C$ from elementary unitaries, also referred to as {\sl gates}, are visualised by means of circuit diagrams. On a more practical level, a quantum algorithm is implemented by designing the arrangement of the elementary gates in a quantum~SDK such as Qiskit~\cite{Qiskit}, Cirq~\cite{cirq}, PyQuil~\cite{smith2016practical} and Pennylane~\cite{bergholm2018pennylane}, see also the textbooks~\cite{ikeAndMike,kaye2006introduction,rieffel2011quantum,wong2022introduction} for known quantum algorithms and their circuit implementations. 

The state vector~$\ket {q(T)}$ at the completion of the computation is interrogated by~\emph{measurement}. According to the measurement postulate of quantum mechanics, the magnitude of the amplitude~$|q_i|^2$ is equal to the probability to find the quantum system in the state labelled~$\ket i$, see~\cite{sakurai1995modern}. After the state is~\emph{collapsed} into the state~$\ket i$, repeated measurements will yield the same result~$\ket i$. Therefore, in quantum computing we can only infer the statistics of components of~$\ket{q(T)}$ or, more generally, the statistics of the projection of~$\ket{q(T)}$ to some subspace. 

In sum, it is important to bear in mind that quantum computing relies exclusively on unitary operations and measurement. Therefore, in applications to computational mechanics it becomes necessary to revisit and reassess existing approaches for their suitability and develop new methods conforming specifically to the quantum computing paradigm. 

\subsection{Opportunities for computational solid mechanics.}
In this paper, we specifically aim to elucidate the suitability---and opportunity---of quantum computing for accelerating, and finally making feasible as a matter of course, concurrent multiscale calculations in solid mechanics. These calculations are concerned with problems characterised by three well-separated scales: The structural or macromechanical scale; the material point or mesomechanical scale; and the subgrid or micromechanical scale. The macromechanical scale encompasses the entire structure or device. It is represented by solid mechanics and discretised, usually, by finite elements. For the materials of interest, the constitutive behaviour of the corresponding Gauss or material points represents the average response of a \emph{representative volume element} (RVE) of material on a mesoscopic scale that is intermediate between the scale of the structure and the material microstructure. The average behaviour of the RVE is in turn the result of a stand-alone calculation that resolves the material macrostructure and encodes the---supposedly known---material behaviour at the microscale. A veritable bestiary of computational schemes, such as sequential microsctructures \cite{Conti:2002, Aubry:2003, Conti:2007, Hansen:2010,cirak2014computational}, computational homogenisation \cite{Geers:2010, segurado:2018}, the Fast-Fourier Transform (FFT) \cite{moulinec1998numerical, gierden2022review}, FE${}^2$ \cite{Smit:1998, Feyel:1999}, discrete dislocation dynamics \cite{arsenlis:2007}, mixed continuum-atomistic methods \cite{ortiz:2001, miller:2007} and data-driven approaches \cite{Leygue:2018, Karapiperis:2021, Korzeniowski:2021, Liu:2022, herath2022computational, Weinberg:2023}, among others, have been developed in order to evaluate RVE behaviour. However, the on-the-fly evaluation of the RVEs concurrently with the structural finite-element calculation is exceedingly costly and frequently beyond the scope of even the largest computational platforms, which motivates the need for a radical paradigm shift such as quantum computing.

\subsection{Related work}
The development of quantum techniques for computational mechanics is still in its infancy. There are several foundational quantum computing algorithms, as reviewed in~\cite{ adedoyin2018quantum,lin2022lecture,tosti2022review}, that will likely play a key role in quantum computational mechanics. The quantum linear system algorithm (QLSA) proposed by Harrow, Hassidim and Lloyd~\cite{harrow2009quantum} aims to solve sparse linear systems of equations~$Av = f$ with~$A \in \mathbb C^{N\times N}$ and~$v,f \in \mathbb C^N$ with a computational complexity~$O(\log(N) s^2 \kappa^2/ \epsilon)$. Here,~$s$ denotes the bandwidth, $\kappa$ the condition number and~$\epsilon$ the targeted accuracy. The complexity of the QLSA algorithm can been further improved in terms of its dependency on the condition number and accuracy~\cite{ambainis2012variable} and~\cite{childs2017quantum}, respectively. The complexity estimates for QLSA assume that the components of~$A$ and~$f$ are already amplitude--encoded. In practice, QLSA cannot perform better than linear if~$A$ and~$f$ are arbitrary and must be first encoded using state preparation. The amplitude encoding of~$A$ can be circumvented by discretising the computational mechanics problem directly on a quantum computer, eliminating the need for its state preparation. The direct quantum discretisation of Poisson problems on equidistant grids, in combination with QLSA, has been considered, e.~g., in~\cite{cao2013quantum, childs2021high,vazquez2022enhancing}.   The solution of non-stationary transient equations, i.~e., heat and wave equations, has been considered as well~\cite{linden2022quantum, costa2019quantum}. For amplitude encoding of~$f$ there are exact and approximate algorithms with better than linear complexity if~$f$ has a certain structure or the number of the used ancilla (helper) qubits is unlimited~\cite{araujo2021divide,vazquez2022enhancing, bharadwaj2023hybrid}.

The approaches discussed so far are intended for gate-based quantum computing, which, as mentioned above, is a universal model of quantum computing. Quantum annealing provides an alternative approach for solving linear systems of equations or, more precisely, optimisation problems~\cite{hauke2020perspectives,mohseni2022ising}. In quantum annealing, the solution of a quadratic unconstrained binary optimisation (QUBO) problem is re-expressed as the ground state, i.~e., the eigenstate with the lowest eigenvalue, of a quantum Ising problem. The ground state is determined by letting the quantum system evolve from a known state using special-purpose quantum annealing hardware, such as the D-Wave machine. Quantum annealing has been adapted for the solution of linear systems of equations resulting from finite element discretisation~\cite{srivastava2019box,raisuddin2022feqa,endo2022phase}. It is currently unclear whether quantum annealing can provide an actual speed-up in comparison to classical computing~\cite{ hauke2020perspectives}. 

\subsection{Our contribution.}
We stipulate that a possible way in which quantum computing could insert itself into this framework could be by accelerating the RVE calculations to the point that concurrent multiscale computing becomes feasible in practical times. To test this proposition, we specifically consider the homogenisation of microstructured materials, such as multiphase composites, consisting of materials with different properties or polycrystalline materials. The characteristic length scale of the microstructure is much smaller than the size of the macroscopic specimen to be analysed. Therefore, it is reasonable to derive the effective material properties, such as diffusivity or Young's modulus, of the homogenised macroscopic boundary value problem from subgrid micro-scale boundary value problems fully considering the microstructure. We specifically consider cubic RVEs subject to periodic boundary conditions and a prescribed average strain. We solve the RVE problem by recourse to the FFT, as pioneered by Moulinec and Suquet~\cite{moulinec1998numerical}, see also the recent review paper~\cite{gierden2022review}. 

The FFT has computational complexity~$\mathcal{O} (N \log N)$, where~$N$ is the number of degrees of freedom, which in practice severely limits the size of the RVE. Remarkably, the corresponding quantum algorithm, the Quantum Fourier Transform (QFT), has polylog complexity~$\mathcal{O}((\log N)^2)$, i.~e., a quadratic polynomial in~$\log N$, see~\cite{coppersmith:1994}, which supplies the sought acceleration. However, the development of an efficient RVE solver on a quantum computer raises a number of challenges. Beyond the QFT algorithm and the already-mentioned state preparation, the quantum RVE solver requires the implementation of simple algebraic operations, such as division by a scalar, in Fourier space. Our proposed strategy is to first approximate functions by piecewise polynomials using Chebyshev interpolation and subsequently encode the polynomials utilising a sequence of elementary rotation gates~\cite{woerner2019quantum, stamatopoulos2020option}. Furthermore, we propose a quantum version of the fixed-point iteration designed to minimise the number of costly measurement operations. Finally, we present a number of numerical test cases that bear out the remarkable exponential acceleration characteristic of quantum computing.

%
\section{Essentials of quantum computing \label{sec:background}} 
%
In this Section, we present a brief summary of the foundations of quantum information processing covering aspects of quantum mechanics and computing. For a more comprehensive review we refer to textbooks~\cite{ikeAndMike, kaye2006introduction, rieffel2011quantum, wong2022introduction} and tutorial paper~\cite{adedoyin2018quantum}. 
%
\subsection{Notation and definitions}
%
We use both the Dirac notation from quantum mechanics, also known as bra-ket notation, and matrix notation from linear algebra. As is evident from Table~\ref{tab:notation}, the two notations closely mirror each other. 
\begin{table}[h!]
\setlength{\extrarowheight}{5pt}
\begin{tabular}{l|l|l|l}
                             & Matrix notation & Dirac notation & Examples / Notes \\ \hline
Scalar            & $z \in \mathbb C$        &    $z \in \mathbb C$   &   $z = 1+i$    \\
\rowcolor[HTML]{ECF4FF} 
Complex conjugate of $z$ &   $z^* \in \mathbb C$  &   $z^* \in \mathbb C$ & $(1+i)^*=(1-i)$   \\
Modulus of $z$    &  $|z| \in \mathbb R$       &    $|z| \in \mathbb R$  & $ |z| = \sqrt{z^* z} = \sqrt{ z z^*} $   \\
\rowcolor[HTML]{ECF4FF} 
Vector (or, a ket)     &  $\vec q \in \mathbb C^N$       &    $\ket q \in \mathbb C^N$  & \begin{tabular}[c]{@{}l@{}} Components: $q_j$ \\ $j=\{0, \, 1, \, \dotsc, \, N-1 \}$\end{tabular}    \\
Dual vector (or, the bra) of $\vec q $  &  $\vec q^\dagger \in \mathbb C^N$ &     $\bra q \in \mathbb C^N$ & $\vec q^\dagger = (\vec q^\trans)^* $ \\  
 \rowcolor[HTML]{ECF4FF} 
Inner product between $\vec q$ and $\vec r$  &  $\vec q \cdot \vec r$ &     $ \braket{ q | r} $ & $\vec q \cdot \vec r = \vec q^\dagger \vec r $ \\  
Kronecker (tensor) product     &  $\vec q \otimes \vec r \in \mathbb C^{N \times N} $  &  $\ket q \otimes \ket r \in \mathbb C^{N \times N}$   &   $ \ket q \otimes \ket r  \equiv  \ket q \ket r \equiv  \ket {qr} $    \\
\rowcolor[HTML]{ECF4FF} 
Matrix    &  $\vec A \in \mathbb C^{N \times N}$       &    $A \in \mathbb C^{N \times N}$ &  \begin{tabular}[c]{@{}l@{}} Components: $A_{jk}$ \\ $j,k=\{0, \, 1, \, \dotsc, \, N-1 \}$\end{tabular}     \\
Hermitian conjugate of $A$   &  $\vec A^\dagger \in \mathbb C^{N \times N}$       &    $A^\dagger \in \mathbb C^{N \times N}$ & $\vec A^\dagger = ( A^\trans)^* = ( A^*)^\trans $     \\
\rowcolor[HTML]{ECF4FF} 
Inner product between $\vec q$ and $\vec A \vec r$ & $ \vec q^\dagger \vec A \vec r $ & $\braket{q | A | r}$ & $\braket{q | A | r}^* = \braket{r | A | q}$  \\ 
 \begin{tabular}[c]{@{}l@{}} Standard basis \\ (or, the computational basis) \end{tabular}  & $ \vec e_0, \, \vec e_1, \, \dotsc , \, \vec e_{N-1} $& $ \ket 0, \, \ket 1, \, \dotsc , \, \ket{N-1} $ &
\begin{tabular}[c]{@{}l@{}} $ \vec e_0 = ( 1 \; \;  0 \; \; \dotsc \; \; 0 )^\trans $ \\ $ \vec e_1 = ( 0 \; \;  1 \; \; \dotsc \; \; 0 )^\trans  $  \end{tabular}
\end{tabular}
\caption{Summary of the definitions and notation used in this paper. \label{tab:notation}}
\end{table}
%

%
 \subsection{Quantum systems}
%
In quantum computing, data is encoded via the state of a quantum system consisting of one or more qubits, i.~e., two-state quantum particles. The quantum state of a system of qubits obeys the fundamental laws of quantum mechanics.

%
 \subsubsection{One-qubit systems}
%
The quantum state~$\ket q \in \mathbb C^2$ of a single qubit is expressed as the superposition, or linear combination, of two (pure) states labelled as~\mbox{$\{ \ket{0} =  (1 \, \; 0 )^\trans , \, \ket{1}  = (0 \, \; 1 )^\trans  \}$}, i.~e.,  
\begin{equation} \label{eq:singleQ}
	\ket{q} = q_0 \ket{0} + q_1 \ket{1} = q_0 \begin{pmatrix} 1 \\ 0 \end{pmatrix} + q_1 \begin{pmatrix} 0 \\ 1 \end{pmatrix}  = \begin{pmatrix} q_0 \\ q_1 \end{pmatrix} \, ,
\end{equation}
where $q_0, q_1 \in \mathbb C$ are two coefficients called amplitudes. For comparison, the same state~$\ket q$ expressed in matrix notation reads 
\begin{equation}
	\vec q = q_0 \, \vec e_0 + q_1 \, \vec e_1 = q_0 \begin{pmatrix} 1 \\ 0 \end{pmatrix} + q_1 \begin{pmatrix} 0 \\ 1 \end{pmatrix}  = \begin{pmatrix} q_0 \\ q_1 \end{pmatrix} \, . 
\end{equation}
The kets~$\ket 0$ and~$\ket 1$, or~$\vec e_0$ and~$\vec e_1$, are two basis vectors which are referred to as the \emph{computational basis}.

The state vector~$\ket q$ expressed in a different orthonormal basis (spanning the same Hilbert space~$\mathbb C^2$) will have different coefficients. Assuming that the measuring device can measure the two states~$\{\ket 0, \, \ket 1 \} $, the qubit will be found either in state~$\ket 0$ or~$\ket{1} $ when observed. In quantum mechanics terminology, the state~$\ket q$ will \emph{collapse} either into state~$\ket 0$ or~$\ket 1$. According to the postulates of quantum mechanics, specifically the \emph{Born rule}, the probability~\mbox{$p(\ket 0)$} to observe~$\ket 0$ is 
\begin{equation} \label{eq:probK0}
	p( \ket 0) = | \braket{q| 0} |^2= | q_0^* \braket{ 0 | 0} + q_1^* \braket{ 1 | 0} |^2 =  | q_0^* |^2 = | q_0 |^2 \, .
\end{equation}
Note that~$\bra q = q_0^* \bra 0 + q_1^* \bra 1$ according to Table~\ref{tab:notation}. Similarly, the probability~\mbox{$p(\ket 1)$} to observe~$\ket 1$ is~$p( \ket 1) = |q_1|^2$. The two probabilities must satisfy~$|q_1|^2 + |q_2|^2 = 1$. Once the qubit has been observed, i.~e., the state has collapsed, it will maintain its observed state and repeated measurements will yield the same result. 

As will be detailed in Section~\ref{sec:measurement}, we note that the two states~$\ket 0$ and $\ket 1$ are a property of the measuring device rather than the quantum system. For instance, a measuring device that can measure only the orthogonal states~$\{ \ket r, \, \ket{r^\perp} \}$ will find the qubit in one these two states. The respective probabilities are determined by replacing~$\ket 0$ by $\ket r$ or $\ket {r^\perp}$ in~\eqref{eq:probK0}. Hence, the representation of the intrinsic qubit state~$\ket q$ in a specific basis is a matter of choice by the observer. 

%
 \subsubsection{Multi-qubit systems \label{sec:multQubitSys}}
%
We begin by considering a two-qubit system ($n=2$) with state vector~$\ket q \in \mathbb C^{2^n} \equiv \mathbb C^4$. The elements of the respective $4$-dimensional basis are given by the Kronecker product of the basis states~$\{ \ket{0}, \, \ket{1} \}$ for each of the qubits, namely, 
\begin{equation} 
	\{ \ket 0 \otimes \ket 0 \,  , \quad  \ket 0 \otimes \ket 1  \, ,  \quad \ket 1 \otimes \ket 0  \, , \quad  \ket 1 \otimes \ket 1 \} \, . 
\end{equation}
These four basis states are often abbreviated as 
\begin{equation}
	\ket 0 \otimes \ket 0 \equiv \ket 0 \ket 0 \equiv \ket{00} \, , \quad \ket 0 \otimes \ket 1 \equiv \ket 0 \ket 1 \equiv \ket{01} \, , \quad \ket 1 \otimes \ket 0 \equiv \ket 1 \ket 0 \equiv \ket{10} \, , \quad \ket 1 \otimes \ket 1 \equiv \ket 1 \ket 1 \equiv \ket{11} \, .
\end{equation}
Converting the four binary numbers $00$, $01$, $10$ and $11$ to the decimals $0$, $1$, $2$ and $3$, respectively, yields an alternative labelling for the same states
\begin{equation}
	\ket 0 \equiv \ket 0 \otimes \ket 0 \, , \quad \ket 1 \equiv \ket 0 \otimes \ket 1 \, , \quad \ket 2 \equiv \ket 1 \otimes \ket 0 \, , \quad \ket 3 \equiv \ket 1 \otimes \ket 1 \, .
\end{equation}
After evaluating the Kronecker products, we obtain
\begin{equation}
	 \ket 0 = \begin{pmatrix} 1 \begin{pmatrix}1\\ 0 \end{pmatrix} \\[1.em] 0 \begin{pmatrix} 1\\ 0 \end{pmatrix} \end{pmatrix} = \begin{pmatrix} 1\\ 0 \\ 0\\ 0 \end{pmatrix} , \quad 
	 \ket 1 = \begin{pmatrix} 1 \begin{pmatrix} 0 \\ 1 \end{pmatrix} \\[1.em] 0 \begin{pmatrix} 0\\ 1 \end{pmatrix} \end{pmatrix} = \begin{pmatrix} 0\\ 1 \\ 0\\ 0 \end{pmatrix} , \quad
	 \ket 2 = \begin{pmatrix} 0 \begin{pmatrix}1\\ 0\end{pmatrix} \\[1.em] 1 \begin{pmatrix} 1\\ 0 \end{pmatrix} \end{pmatrix} = \begin{pmatrix} 0\\ 0 \\ 1\\ 0 \end{pmatrix} , \quad
	 \ket 3 = \begin{pmatrix} 0 \begin{pmatrix}0 \\ 1 \end{pmatrix} \\[1.em] 1 \begin{pmatrix} 0\\ 1 \end{pmatrix} \end{pmatrix} = \begin{pmatrix} 0\\ 0 \\ 0\\ 1 \end{pmatrix} \, .
\end{equation}
Using these basis vectors, the state $\ket q \in \mathbb C^4$ of a two-qubit system has the representation
\begin{equation} \label{eq:twoQstate}
	\ket q = \sum_{k=0}^{3} q_k \ket k =  \begin{pmatrix} q_0 \\ q_1 \\ q_2 \\ q_3 \end{pmatrix} \, , 
\end{equation}
with coefficients~$q_k \in \mathbb C$. For $\ket q$ to represent a quantum state of a two-qubit system, the coefficients~$q_k$ must be normalised. The probability to observe the qubit in state~$\ket k$ is~$p(\ket k) = |q_k|^2$. 
 
It bears emphasis that the Kronecker products of two one-qubit quantum states do not span all possible quantum states of the two-qubit state~\eqref{eq:twoQstate}. The Kronecker product of two one-qubit states~$\ket r$ and~$\ket s$ yields the state
\begin{equation} \label{eq:stateSeparable}
 	\ket r \otimes \ket s = \left ( r_0 \ket 0 + r_1 \ket 1 \right ) \otimes \left ( s_0 \ket 0 + s_1 \ket 1 \right ) = 
	\begin{pmatrix} r_0 s_0 \\ r_0 s_1 \\ r_1 s_0 \\ r_1 s_1 \end{pmatrix} \, .
\end{equation}	
For instance, consider a two-qubit system in the state
%
\begin{equation} \label{eq:bellstate}
	\ket q = \frac{1}{\sqrt 2} \ket 0 \otimes \ket 0 + \frac{1}{\sqrt 2} \ket 1 \otimes \ket 1 = 	\begin{pmatrix} 1/\sqrt 2 \\ 0 \\ 0 \\ 1/\sqrt 2 \end{pmatrix} \, .
\end{equation}
Matching (\ref{eq:stateSeparable}) to (\ref{eq:bellstate}) 
requires $r_0 s_0 = r_1s_1 = 1/\sqrt{2}$, but at the same time $r_0 s_1 =0$ and $r_1 s_0 =0$, which has no solutions and shows that (\ref{eq:bellstate}) cannot be represented a Kronecker product of one-qubit states. The states that cannot be represented as the Kronecker product of one-qubit states are referred to as \emph{entangled states}. The non-entangled states are the~\emph{separable states} or \emph{product states}.

Similar to the two-qubit case, the state~$\ket q \in \mathbb C^{2^n}$ of a system of $n$ qubits is given by
\begin{equation} \label{eq:multQstate}
	\ket q = \sum_{k = 0}^{2^n -1} q_k \ket k = \begin{pmatrix} q_0 \\ q_1 \\ \vdots \\ q_{2^n-1} \end{pmatrix} \, , 
\end{equation}
where~$q_k \in \,\mathbb C$. The index~$k$ is frequently expressed as a binary
\begin{equation} \label{eq:multiQlabeling}
	k = k_{0} 2^{n-1} + k_{1} 2^{n-2} + \dotsc +  k_{n-1} 2^0 \equiv  k_{0} k_{1} \dotsc  k_{n-1} \, ,
\end{equation}
so that~\eqref{eq:multQstate} can be written as 
\begin{equation}
	\ket q = \sum_{k = 0}^{2^n -1} q_k \ket k = \sum_{k_{0}=0}^{1} \sum_{k_{1} =0}^{1} \dotsc \sum_{k_{n-1}=0}^{1} q_{k_{0} k_{1} \dotsc   k_{n-1}} \ket{k_{0} k_{1} \dotsc  k_{n-1}}  \, .
\end{equation}
Moreover, similarly to two-qubit states, each of the states~$\ket k$ is obtained as the Kronecker product of single-qubit states, namely, 
\begin{equation}
	  \ket{k_{0} } \otimes \ket{ k_{1} } \otimes \dotsc \otimes \ket{ k_{n-1}} \equiv \ket{k_{0} } \ket{ k_{1} } \dotsc  \ket{ k_{n-1}} \equiv	\ket{k_{0} k_{1} \dotsc  k_{n-1}} \, .
\end{equation}

The inevitability of entangled states is even more apparent in case of~$n>2$. Each separate qubit has two coefficients so that a separable state with $n$ qubits can only have up to~$2n$ distinct coefficients, i.~e., the separable state spans a space of dimension $2n$. By contrast, an entangled state can have up to~$2^n$ distinct coefficients, which spans the entire space of quantum states of dimension $2^n$. We note that the entanglement property of quantum systems is, in addition to their superposition property, fundamentally different from classical systems such as a system of $n$ binary coins with the states head and tail~\cite{jozsa2003role}. 
%
 \subsection{Quantum gates and circuits \label{sec:gatesCircuits}}
%
At the most abstract level, a quantum computer maps a state~$\ket q$ to a new state~\mbox{$U \ket q$} for a given unitary matrix~$U$. The coefficients of the state~$\ket q$ are the input and the coefficients of~\mbox{$U \ket q$} are the output. The unitarity requirement,~$U^\dagger = U^{-1}$, is a postulate of quantum mechanics and ensures that the mapped state remains normalised. 

In gate-based quantum computing, the unitary matrix~$U$ is composed of a sequence of smaller elementary unitary matrices acting on only one or a few qubits at a time. This approach is possible because the multiplication and Kronecker products of unitary matrices are also unitary. In the \emph{quantum circuit}, the elementary unitary matrices correspond to~\emph{gates} and we therefore refer to elementary matrices interchangeably as gates. This section summarises the most widely used gates and discusses how they are combined into quantum circuits implementing a unitary transformation~$U$. 


%
 \subsubsection{One-qubit gates}
%
The three most common one-qubit gates are the Pauli gates and the unitary gate,
\begin{equation}
	X = \begin{pmatrix*}[r]
		0 & 1 \\ 
		1 & 0
		\end{pmatrix*}
		\, , 
		\quad 
	Y = \begin{pmatrix*}[r]
		0 & -i \\ 
		i & 0
		\end{pmatrix*}
		\, , 
		\quad 
	Z = \begin{pmatrix*}[r]
		1 & 0 \\ 
		0 & -1
		\end{pmatrix*}
		\, ,	
		\quad 
	I = \begin{pmatrix*}[r]
		1 & 0 \\ 
		0 & 1
		\end{pmatrix*} ,
\end{equation}
respectively.
%
%
%
The Pauli matrices are Hermitian and involutory (self inverse). Taking the  matrix exponential of the Pauli matrices yields the rotation gates
%
%
\begin{equation} \label{eq:rotationGates}
	R_X(\theta) = e^{-i \theta X /2 } = 
	\begin{pmatrix*}[r]
		\cos \frac{\theta}{2} & - i \sin \frac{\theta}{2} \\[0.2em]  - i \sin \frac{\theta}{2} & \cos \frac{\theta}{2} 
	\end{pmatrix*} \, , \; \; 
		R_Y(\theta) = e^{-i \theta Y /2 } =
	\begin{pmatrix*}[r]
		\cos \frac{\theta}{2} & - \sin \frac{\theta}{2} \\[0.2em]  \sin \frac{\theta}{2} & \cos \frac{\theta}{2} 
	\end{pmatrix*} \, , \; \;
		R_Z(\theta) = e^{-i \theta Z /2 }  =
	\begin{pmatrix*}[c]
		e^{-i \theta/2} & 0 \\[0.2em]  0 & e^{i \theta/2} 
	\end{pmatrix*} \, .
\end{equation}
%
%
Other widely used one-qubit gates include the Hadamard and the phase gates, 
\begin{equation} \label{eq:hadamard_phase}
	H = \frac{1}{\sqrt 2} \begin{pmatrix*}[r]
		1 & 1 \\ 
		1 & -1
		\end{pmatrix*} \, , \quad 
	P(\theta) = \begin{pmatrix*}[c] 
		1 & 0 \\ 
		0 & e^{i \theta}
	\end{pmatrix*} \, ,
\end{equation}
respectively.
%
 \subsubsection{Multi-qubit gates} \label{sec:C-NOT}
%
We first consider gates defined as Kronecker products of one-qubit gates. Note that the Kronecker product of two or more unitary matrices is a unitary as well, as required. By way of example, consider the two-qubit gate
\begin{equation}
	X \otimes Y = 
 	\begin{pmatrix*}[r]
		0 & 1 \\ 
		1 & 0
		\end{pmatrix*}
		\otimes 
	\begin{pmatrix*}[r]
		0 & -i \\ 
		i & 0
		\end{pmatrix*}
		= 
	\begin{pmatrix*}[r]
		0
		\begin{pmatrix*}[r] 
			0 & -i \\ 
			i & 0
		\end{pmatrix*} 
		&
		1
		\begin{pmatrix*}[r] 
			0 & -i \\ 
			i & 0
		\end{pmatrix*} 
		\\[1.em] 
		1
		\begin{pmatrix*}[r] 
			0 & -i \\ 
			i & 0
		\end{pmatrix*} 
		&
		0
		\begin{pmatrix*}[r] 
			0 & -i \\ 
			i & 0
		\end{pmatrix*} 
	\end{pmatrix*}
		=
	\begin{pmatrix*}[r]
		0 & 0 & 0 & -i \\
		0 & 0 & i & 0 \\
		0 & -i & 0 & 0 \\
		i & 0 & 0 & 0
	\end{pmatrix*}
	\, .
\end{equation}
If this gate is applied to a non-entangled product state~$ \ket r \otimes \ket s$ it is equivalent to applying first~$X$ to~$\ket r$ and~$Y$ to~$\ket s$ and then taking their Kronecker product, i.~e.,  
\begin{equation}
 	(X \otimes Y) (\ket r \otimes \ket s) =  (X \ket r ) \otimes ( Y \ket s) \, .
\end{equation}
Hence, starting with a product multi-qubit state applying only single qubit gates the state will remain a non-entangled product state. 

To achieve entanglement a $CNOT$, also called a controlled-NOT or controlled-$X$ gate, is required. The~$CNOT$ gate is defined as 
\begin{equation} \label{eq:cnot}
	CNOT = \ket 0 \bra 0 \otimes I + \ket 1\bra 1 \otimes X \, ,
\end{equation}
and has the matrix representation
\begin{equation} \label{eq:cnotMatrix}
	CNOT = \begin{pmatrix} 1 & 0 \\ 0 & 0\end{pmatrix} \otimes \begin{pmatrix} 1 & 0 \\ 0 &1\end{pmatrix} + 
	\begin{pmatrix} 0 & 0 \\ 0 & 1\end{pmatrix} \otimes \begin{pmatrix} 0 & 1 \\ 1 &0 \end{pmatrix} =  \begin{pmatrix} 1 & 0 & 0 & 0 \\ 0 & 1 & 0 & 0 \\ 0 & 0 & 0 & 1 \\ 0 & 0 & 1 & 0\end{pmatrix} \, .
\end{equation}
In studying the action of the~$CNOT$ gate, or any other gate for that matter, on a state 
\begin{equation}
	\ket q = \sum_{k=0}^3 q_k \ket k = \sum_{k_0=0}^1 \sum_{k_1 = 0}^1 q_{k_0 k_1} \ket{k_0 k_1} \, ,
\end{equation} 
it is sufficient to focus on its action on a single basis element~\mbox{$\ket k = \ket{k_0} \otimes \ket{k_1} \equiv  \ket{k_0 k_1} $}. This is possible because of the linearity of quantum transformations. According to~\eqref{eq:cnot} and~\eqref{eq:cnotMatrix}, the~$CNOT$ gate maps the four basis elements as follows:
\begin{equation}
	CNOT \colon \ket {0 0} \mapsto \ket {0  0} ; \quad \ket {0 1} \mapsto \ket {0 1} ; \quad \ket {1 0} \mapsto \ket {1 1} ; \quad \ket {1  1} \mapsto \ket {1  0} .
\end{equation}
Here, the left qubit $k_0$ is the \emph{control} and the right qubit~$k_1$ is the~\emph{target}. Evidently, the action of the~$CNOT$ gate is to flip the target qubit~$k_1$ if the control qubit~$k_0$ is in state~$\ket 1$; and to leave it unchanged otherwise. 

In passing, we note to obtain the controlled two-qubit versions of other single-qubit gates it is sufficient to replace~$X$ gate in~\eqref{eq:cnot} with the respective single qubit gate.

%
 \subsubsection{Quantum circuits}
%
As stated earlier, quantum computation can be summarised as the mapping of an $n$-qubit state $\ket q \in \mathbb C^{2^n}$ into the new state $U \ket q \in \mathbb C^{2^n}$ for the purpose of doing some desired computation. The construction of the unitary matrix $U \in \mathbb C^{2^n\times 2^n}$ out of one-qubit and multi-qubit gates is commonly visualised using \emph{circuit diagrams}~\cite{feynman1986quantum}. 

\begin{figure}
	\centering
 	\subfloat[][\label{fig:scirca}] {	
	\scalebox{1.0}{
	\Qcircuit @C=1.1em @R=1.em @!R { \\
	 	 \lstick{ \ket{{k}_{0}} = \ket{{0}} } & \gate{\mathrm{I}} & \qw & \qw\\
	 	 \lstick{ \ket{{k}_{1}} = \ket{{0}} } & \gate{\mathrm{X}} & \qw & \qw\\
	}}
		} \hspace{0.2\textwidth}
	\subfloat[][\label{fig:scircb}] {			
	\scalebox{1.0}{
	\Qcircuit @C=1.1em @R=1.em @!R { \\
	 	 \lstick{\ket{{k}_{0}} = \ket{{0}} } & \qw \qw & \qw & \qw\\
	 	 \lstick{\ket{{k}_{1}} = \ket{{0}} } & 
  \gate{\mathrm{X}} &  \qw & \qw \\
 }}
	}
\caption{A basic quantum circuit with two qubits $k_0$ and $k_1$. As indicated in (b) the identity gate~$I$ is usually omitted in circuit diagrams so that circuits (a) and (b) are equivalent. The states of both qubits are initialised with $\ket {k_0} = \ket 0$ and $\ket {k_1} = \ket 0$ so that \mbox{$\ket q = \ket {k_0} \otimes \ket {k_1} = \ket 0 \otimes \ket 0$} which is abbreviated as~\mbox{$ \ket q = \ket{0}\ket{0}$} or~\mbox{$ \ket q = \ket{00}$}. The output of the circuit is~\mbox{$ (I \otimes X) (\ket 0 \otimes \ket 0) = I \ket 0 \otimes X \ket 0= \ket 0 \otimes \ket 1 \equiv \ket 0 \ket 1 \equiv \ket {01}$.} 
\label{fig:scirc}}
\end{figure}
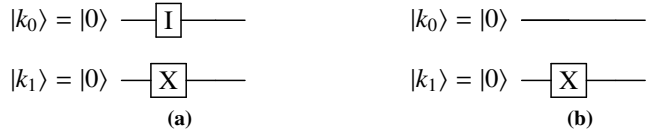
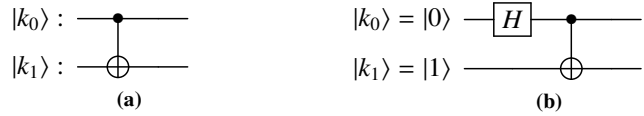
\begin{figure}
  \centering
	\subfloat[][\label{fig:tcirc_a}] {	
	\centering 
	\scalebox{1.0}{
	\Qcircuit @C=1.1em @R=1.em @!R { \\
	 	\lstick{\ket{{k}_{0}} : } &\ctrl{1} & \qw & \qw\\
	 	\lstick{\ket{{k}_{1}} : } & \targ & \qw & \qw\\
	 }}
	}  \hspace{0.2\textwidth}
	\subfloat[][\label{fig:tcirc_b}] {	
	\centering 
\scalebox{1.0}{
\Qcircuit @C=1.1em @R=0.6em @!R { \\
	 	\lstick{ \ket{{k}_{0}} = \ket{{0}} } & \gate{H} & \ctrl{1} & \qw & \qw\\
	 	\lstick{ \ket{{k}_{1}} = \ket{{1}} } & \qw & \targ & \qw & \qw\\
}}
}
	\caption{Quantum circuits with two qubits~$k_0$ and $k_1$. The circuit (a) depicts the~$CNOT$ gate where $k_0$ is the control and $k_1$ the target wire. The respective unitary matrix is given in~\eqref{eq:cnotMatrix}. The state of the target~$k_1$ is flipped only when the state of the control~$k_0$ is $\ket 1$, otherwise it remains unchanged. The output of the circuit (b) is the entangled state $\frac{1}{\sqrt 2} (\ket 0 \otimes \ket 0 + \ket 1 \otimes \ket 1) \equiv \frac{1}{\sqrt 2} (\ket {0} \ket{0} + \ket {1} \ket{1}) \equiv \frac{1}{\sqrt 2} (\ket {00} + \ket {11}) $. \label{fig:tcirc}}
\end{figure}

In the circuit diagrams depicted in Figures~\ref{fig:scirc} and~\ref{fig:tcirc}, each line is referred to as a \emph{wire} and represents a qubit. The circuit is read from left to right and shows the sequence of applied single and two-qubit gates. There is no joining or splitting of wires which would violate the unitarity requirement for~$U$. As illustrated in Figure~\ref{fig:scirc}, the identity gates are usually omitted in the diagrams. 


%

In the basic circuit depicted in Figure~\ref{fig:scirc} the computation starts with two qubits~$k_0$ and~$k_1$ both in the state~$\ket 0$. The corresponding state vector is 
\begin{equation}
	\ket q = \ket {k_0} \otimes \ket {k_1} = \ket 0 \otimes \ket 0 = \begin{pmatrix} 1 \\ 0 \\ 0 \\ 0 \end{pmatrix}
\end{equation}
Next, we apply to the qubit $k_1$ the Pauli $X$ gate and to the qubit $k_0$ the identity gate~$I$. The resulting state is given by 
\begin{equation}
	\ket q = (I \otimes X) (\ket {k_0} \otimes \ket {k_1}) = I \ket 0 \otimes X \ket 0 =  \begin{pmatrix} 1 \\ 0 \end{pmatrix} \otimes  \begin{pmatrix} 0 & 1 \\ 1 & 0 \end{pmatrix} \begin{pmatrix} 1 \\ 0 \end{pmatrix} =  \begin{pmatrix} 0 \\ 1 \\ 0 \\ 0 \end{pmatrix} \, .
\end{equation}
We verify that~$U = I \otimes X $ is unitary for entire circuit, as required. 

The entangled state given in equation~\eqref{eq:bellstate} can be obtained with the circuit depicted in Figure~\ref{fig:tcirc_b}. The sequence of the mappings is applied as follows: 
\begin{align}
\begin{split}
	(H \otimes I) & \colon  \ket 0 \otimes \ket 0 
  \mapsto 
  H \ket 0 \otimes I \ket 0 =  \frac{1}{\sqrt 2} \left ( \ket 0 + \ket 1 \right ) \otimes \ket 0   \, , 
   \\ 
	CNOT 
  &\colon  \frac{1}{\sqrt 2} \left ( \ket 0 \otimes \ket 0+ \ket 1 \otimes \ket 0 \right ) 
  \mapsto  
   \frac{1}{\sqrt 2} \left ( \ket 0 \otimes \ket 0 + \ket 1 \otimes \ket 1 \right ) \, . 
\end{split}
\end{align}
For the $CNOT$ the left qubit ($k_0$) is the control qubit and the right qubit ($k_1$) is the target qubit. The target qubit state is
flipped when the control qubit is in state~$\ket 1$.  The mapping by~$CNOT$ can also be obtained directly from the definition~\eqref{eq:cnot} or~\eqref{eq:cnotMatrix} of the~$CNOT$ gate and noting the orthonormality of the states~$\ket 0$ and~$\ket 1$.

%
 \subsection{Measurement \label{sec:measurement}}
In quantum systems the outcome of a measurement is random and measurement has an irreversible effect on the state of a system. When a quantum system with the state vector~$\ket q \in \mathbb C^{2^{n}}$ is measured we will find it in one of its~$2^n$ possible states. As mentioned, according to the Born rule, the probability of observing the system in the specific state~$\ket {k}$ is~$p(\ket{k}) = |q_{k}|^2$ . The measurement process collapses the quantum state so that after measurement the state vector becomes~$ \ket q = \ket{k}$. Consequently, we can infer the probabilities~$ \{ p(\ket 0), \,  p(\ket 1), \dotsc, \,  p( \ket{2^n-1}) \}$ only by considering the statistics of several quantum computations with the same circuit. 

Evidently, a state vector~$\ket q$ can be expressed in different bases so that the possible outcomes of a measurement depend on the specific basis imposed by the measurement apparatus. This realisation leads to the notion of projective measurement. 
%
A complete set of orthogonal projectors~$P_j$ with the properties~$\sum_j P_j = I$ and~$P_j P_{k} = \delta_{jk}$ can be, for instance, obtained from an orthonormal basis~$\{ \ket {r_0}, \, \dotsc, \, \ket{r_{2^{n} -1}}\}$ as
\begin{equation}
	P_j = \ket{ r_j } \bra{r_j} \ . 
\end{equation}
%
%
In projective measurement, the probability of observing the system in the  state~$\ket {r_j}$ is
\begin{equation}
	p(\ket{r_j}) = | \braket {r_j | q} |^2 = \braket{r_j | q}^* \braket{r_j | q} = \braket{q|r_j} \braket{r_j | q} = \braket{q | P_j | q } \, .
\end{equation}
Note that the probabilities for all~$2^n$ states add up to one as desired. Furthermore, during the measurement the system collapses to the state 
\begin{equation}
	\frac{P_j \ket q}{ | \braket{r_j | q} |} 
  = 
  \frac{P_j \ket q}{\sqrt{p(\ket{r_j})}} \, .
\end{equation}

Projective measurement can be used to derive the rules for partial measurement. For instance, in a two qubit system the two projectors 
\begin{equation}
	P_0 = I \otimes \ket 0 \bra 0\, , \quad P_1 = I \otimes \ket 1 \bra 1 \, , 
\end{equation}
form a complete set. They measure the probability of finding the right qubit in the state~$\ket 0$ or $\ket 1$, respectively.  Hence, applying~$P_0$ to a quantum state~$\ket q = \frac{1}{\sqrt 2} (\ket 0 \ket 0 + \ket 1 \ket 1 )  $, obtained by the circuit depicted in Figure~\ref{fig:tcirc}b, yields the probability
\begin{equation}
	p( \ket 0) = \braket{ q  | P_0 | q } = \frac{1}{2} \, ,
\end{equation}
and the state collapses to 
\begin{equation}
	\frac{P_0 \ket q}{\sqrt{p(\ket 0)}} = \sqrt{2} P_0 \ket q = \ket 0 \ket 0  \, .
\end{equation}

%
\subsection{Quantum algorithm complexity} \label{sec:complexity}
The complexity of a quantum algorithm can be characterised in three main ways: the number of one-qubit gates, the number of two-qubit gates and the depth of the quantum circuit. 
We assess the complexity of our algorithms using the number of quantum gates. 
To this end, we express our quantum circuits using only the two-qubit $CNOT$ gate and the one-qubit generic rotation gate
\begin{equation}
    U_3(\theta, \phi, \lambda) = 
    \begin{pmatrix} \cos\left(\frac{\theta}{2} \right) & -e^{i\lambda}\sin\left( \frac{\theta}{2}\right) \\ e^{i\phi}\sin\left(\frac{\theta}{2} \right) & e^{i(\phi+\lambda)}\cos\left(\frac{\theta}{2}\right) \end{pmatrix} \, ,
\end{equation}
where~$\theta$, $\phi$ and~$\lambda$ are three angles. The gate set $\{ CNOT, \, U_3\}$ is universal in the sense that any quantum circuit can be expressed using its two gates. For instance, it is easy to verify that~\mbox{$U_3(\pi, \pi, \pi/2) = X$} or~\mbox{$U_3(\theta, 0, 0) = R_Y(\theta)$}. Reexpressing a quantum circuit in terms of a given of universal gate set is referred to as \emph{transpiling} and is provided as a standard operation in quantum SDKs such as Qiskit. We further assume an ideal scenario and ignore errors in the quantum gates and the consequent cost of error correction. 

%
\section{Model problem: Computational homogenisation \label{sec:compHomog}} 
%

In the remainder of the paper, we show that---at least---some problems in computational mechanics can be reformulated within the quantum computing framework just outlined and that such reformulation indeed pays off in the form of a game-changing exponential speed-up of the calculations. We demonstrate these points with the aid of a specific example: computational homogenisation. 

For clarity, we restrict attention to computational homogenisation in its simplest form: periodic composites with a scalar solution field governed by a non-homogeneous Laplace equation with average gradient constraints~\cite{toshioMura-1987, moulinec1998numerical, michel1999effective}. Such problems arise, for instance, in the context of electrical and heat conduction, flow in porous media or antiplane elasticity. In computational homogenisation, the solution field is determined by solving concurrently macroscopic and microscopic-level problems. The gradient of the solution field of the macroscopic problem serves as the input to the microstructural problem. In turn, the corresponding macroscopic fluxes are obtained as the averages of the microscopic fluxes, thus closing the micro-macro handshake. 
A complete authoritative account of elliptic homogenisation problems can be found in \cite{Cioranescu:1999}
%
\subsection{Problem formulation}
%
The microscopic problems of interest are concerned with the response of a representative volume element (RVE), cf.~\cite[\S 10.2]{Cioranescu:1999}. We specifically assume an RVE domain in the shape of the square~\mbox{$\Omega \equiv (0, \, L)^2 \in \mathbb R^2 $} with an edge length of~$L$, and the coordinates of the points~$\vec x \in \Omega$ are denoted as~\mbox{$\vec x = ( x_0 \quad x_1)^\trans$}. For definiteness, we consider a model antiplane elasticity problem and name our variables accordingly. The shear modulus~$\mu (\vec x)$ in the RVE is periodic, i.~e., 
\begin{equation}
	\mu(x_0, \, x_1) = \mu (x_0 + L, \, x_1) = \mu (x_0 , \, x_1+L) \, . 
\end{equation}
The RVE is subject to a uniform average strain vector~$\overline {\vec \gamma} \in \mathbb R^2$ handed down by the macroscopic problem. The respective deformation of the RVE is characterised by a scalar transverse displacement field~$u(\vec x) \in \mathbb R$ with shear-strain vector 
\begin{equation} \label{eq:shearStrain}
	\vec \gamma (\vec x) = \nabla u (\vec x) \, ,  
\end{equation}
where~$\nabla$ denotes the gradient operator and~$\vec \gamma (\vec x) \in \mathbb R^2$. The displacement field~$u(\vec x) \in \mathbb R$ can be represented as an affine component matching~$\overline {\vec \gamma}$ and a fluctuating component~$v(\vec x)$, so that 
\begin{equation} \label{eq:dispDecomposition}
	u(\vec x) = \overline{\vec \gamma } \cdot \vec x + v(\vec x) \, , 
\end{equation}
which implies the strain decomposition 
\begin{equation} \label{eq:dispDecomposition}
	\vec \gamma (\vec x) = \overline{\vec \gamma } + \nabla v(\vec x) \, .
\end{equation}
The fluctuating displacement~$v (\vec x) \in \mathbb R$ must satisfy the periodicity condition \cite[\S~2.1]{Cioranescu:1999}
\begin{equation} \label{eq:dispPeriodicitiy}
	v (x_0, \, x_1) = v(x_0 + L, \, x_1) = v(x_0, \, x_1 +L),
\end{equation}
therefore results in the zero-average condition \cite[Thm.~4.26]{Cioranescu:1999}
\begin{equation} \label{eq:dispZeroAverage}
  \frac{1}{L^2} \int_0^L \int_0^L \nabla v(\vec x) \D x_0 \D x_1 = 0 \, .
\end{equation}
The decomposition~\eqref{eq:dispDecomposition}, together with the periodicity and zero-average condition, \eqref{eq:dispPeriodicitiy} and \eqref{eq:dispZeroAverage}, respectively, ensure that~$\overline{\vec \gamma}$ is indeed the average strain. Indeed, we verify that
\begin{equation}
	\frac{1}{L^2} \int_0^L \int_0^L \vec \gamma \D x_0 \D x_1 = \frac{1}{L^2} \int_0^L \int_0^L \left ( \overline{\vec \gamma } + \nabla v(\vec x)  \right ) \D x_0 \D x_1 = \overline{\vec \gamma} \, .
\end{equation}
The periodicity of~$v(\vec x)$ may be enforced simply by appending Dirichlet boundary conditions of the form (cf.~\cite[Prop.~3.49]{Cioranescu:1999})
\begin{subequations}
\begin{align}
  v(x_0, \, 0) = v(x_0, \, L) , 
  & \quad
  x_0 \in (0,L) , 
  \\ 
  v(0, \, x_1) = v(L, \, x_1) ,
  & \quad
  x_1 \in (0,L) .
\end{align}
\end{subequations}
The stress field vector~$\vec \sigma (\vec x) \in \mathbb R^2$ of the RVE satisfies the equilibrium equation 
\begin{align} \label{eq:equilibrium}
	\nabla \cdot \vec \sigma (\vec x) = 0 \, , 	
\end{align}
and the constitutive equation 
\begin{equation} \label{eq:constitutive}
	\vec \sigma(\vec x) = \mu (\vec x) \vec \gamma (\vec x) = \mu (\vec x)  \left ( \overline{\vec \gamma } + \nabla v(\vec x) \right ) \, .
\end{equation}
Hence, the boundary value problem for microstructure can be summarised as 
\begin{equation} \label{eq:homogBVP}
	\nabla \cdot \left ( \mu (\vec x)   \nabla v(\vec x) \right ) + \overline {\vec \gamma} \cdot \nabla \mu (\vec x) = 0  \, ,
\qquad \vec x \in \Omega ,
\end{equation}
subject to the periodicity and zero-average conditions \eqref{eq:dispPeriodicitiy} and \eqref{eq:dispZeroAverage}, respectively.

Due to the non-constant~$\mu(\vec x)$, it is not possible to solve the boundary value problem~\eqref{eq:homogBVP} directly using the Fourier transform. Therefore, following Moulinec and Suquet~\cite{moulinec1998numerical} we introduce a constant reference shear modulus~$\mu^0$ and rewrite the constitutive equation~\eqref{eq:constitutive} as 
\begin{equation} \label{eq:homogBVPlin}
	 \vec \sigma(\vec x) = ( \mu (\vec x) - \mu^0) \vec \gamma(\vec x) + \mu^0 \vec \gamma(\vec x)  = \vec \tau(\vec x) +  \mu^0 \vec \gamma(\vec x)  \, , 
\end{equation}
where~$\vec \tau(\vec x)$ is referred to as the polarisation stress. Next, we use the equilibrium~\eqref{eq:equilibrium} to define the fixed point iteration 
\begin{equation} \label{eq:homogBVP}
	 \mu^0 \nabla \cdot \nabla v^{(s+1)}(\vec x) + \nabla \cdot \vec \tau^{(s)} (\vec x) = 0 \, . 
\end{equation}
The polarisation stress~$\vec \tau^{(s)}$ at iteration step~$s$ depends on the known displacement~$v^{(s)}$ and the applied strain~$\overline{\vec \gamma}$.

%
\subsection{Fourier representation \label{sec:periodicSol}} 
%
In preparation for the Fourier discretisation of the incremental RVE problem, we consider a polarisation stress vector in the form of a monochromatic wave with a period~$L$, angular wave vector~\mbox{$\vec \xi = (\xi_0 ,\, \xi_1 )^\trans$} with~\mbox{$\xi_0 = \xi_1 = 2 \pi / L$} and a complex amplitude~\mbox{$\hat {\vec \tau} \in \mathbb C^2$}, i.~e.,
\begin{equation} \label{eq:tauSingleWave}
	\vec \tau (\vec x) = \hat {\vec \tau} (\vec \xi) e^{i \vec \xi \cdot \vec x} \, ,
\end{equation}
where~$i^2 = -1$ and~$\vec \xi \cdot \vec x = \xi_0 x_0 + \xi_1 x_1$. The corresponding displacement solution is another monochromatic wave with the same wave vector~$\vec \xi$ but a yet unknown amplitude~$\hat v (\vec \xi) \in \mathbb C$, namely, 
\begin{equation} \label{eq:vSingleWave}
	v (\vec x) = \hat v(\vec \xi) e^{i \vec \xi \cdot \vec x} \, .
\end{equation}
Note that the components of~$\vec \tau (\vec x)$ and~$v (\vec x)$ can have different phases given that the amplitudes~$ \hat {\vec \tau} $ and~$\hat v(\vec \xi) $ are complex valued. Introducing the expressions~\eqref{eq:tauSingleWave} and~\eqref{eq:vSingleWave} into~\eqref{eq:homogBVP}, dropping the iteration counter~$s$ for simplicity, yields the displacement amplitude as
\begin{equation}
	- \mu^0 (\vec \xi \cdot \vec \xi ) \hat v (\vec \xi) + i \vec \xi \cdot \hat {\vec \tau } (\vec \xi) = 0 \quad \Rightarrow \quad \hat v (\vec \xi) = \frac{1}{\mu^0} \frac{ i \vec \xi \cdot \hat {\vec \tau } (\vec \xi) }{\vec \xi \cdot \vec \xi } \, .
\end{equation}
Substituting into (\ref{eq:vSingleWave}) gives the full solution of the RVE problem as the wave
\begin{equation} \label{eq:RVEsolution}
	v (\vec x) = \frac{1}{\mu_0} \frac{ i \vec \xi \cdot \hat {\vec \tau } (\vec \xi) }{\vec \xi \cdot \vec \xi } e^{i \vec \xi \cdot \vec x} \, .
\end{equation}
By differentiation, the fluctuation strains finally follow as
\begin{equation}
	\nabla \vec v(\vec x) 
  = 
  -
  \frac{1}{\mu_0} 
  \frac
  {
    \vec \xi \cdot \hat {\vec \tau } (\vec \xi) 
  }
  {
    \vec \xi \cdot \vec \xi 
  } 
  \, \vec \xi \,
  e^{i \vec \xi \cdot \vec x} 
  \equiv
  \hat{\vec \Gamma}(\vec \xi) 
  \cdot 
  \hat{\vec \tau} (\vec \xi) e^{i \vec \xi \cdot \vec x}  \, .	
\end{equation}
%

%
\subsection{Discrete-Fourier Transform (DFT) approximation \label{sec:fourierDiscretization}}
%
We approximate $v(\vec x)$ over the RVE domain~$\Omega$ by recourse to band-limited approximation, also referred to as Whittaker-Shannon interpolation~\cite{trefethen2000spectral, Mallat:2009}. The band-limited approximation of the solution field is defined as
\begin{equation} \label{eq:fourierApprox}
	v^h(\vec x) 
  \approx 
   \frac{1}{N} \sum_{k^0=0}^{N-1} \sum_{k^1=0}^{N-1} \hat v_{\vec k} e^{i \vec \xi_{\vec k} \cdot \vec x }  \, , 
\end{equation}
where $N$ is an even positive integer, 
$\vec k = ( k^0, \, k^1)$ is a multi-index and 
\begin{equation}
	\vec \xi_{\vec k} 
  = 
  \begin{pmatrix}
    \dfrac{2 \pi r(k^0)}{L} & 
    \dfrac{2 \pi r(k^1)}{L} 
  \end{pmatrix}^\trans
  \, ,
\end{equation}
are discrete wave vectors. 
%
%
The function
\begin{equation} \label{eq:componentsToFrequencies}
	r(k) = \begin{cases}
		k & 0 \le k < N/2 \\
		k - N & N/2 \le k < N \, ;
	\end{cases}
\end{equation}
defines a cyclic relabelling of the indices, see~\cite[Ch. 3]{trefethen2000spectral} and~\cite[Ch. 2]{brunton2019data}. 

A straightforward derivation shows that $\hat{v}_{\vec k}$ coincides with the discrete Fourier transform (DFT) of the array \mbox{$v^h(x_{\vec k})$} sampled over the lattice of points
\begin{equation}
  \vec x_{\vec k} 
  = 
  \frac{L}{N} \, \vec k
  \, , 
  \quad k^0 = 0, \, \dotsc , \, N-1 \, , 
  \quad k^1 = 0, \, \dotsc , \, N-1 \, .
\end{equation}
Thus, the multi-index~$\vec k$ can be identified as grid-point labels of the RVE domain discretisation. 
%
%
The DFT and its inverse define unitary operators that can be very efficiently computed on a quantum computer using the QFT algorithm, as detailed in Section~\ref{sec:qft}, 

All periodic fields appearing in the RVE problem~\eqref{eq:homogBVP} can now be approximated using band-limited interpolation~\eqref{eq:fourierApprox}, representing a linear combination, or superposition, of $N^2$ monochromatic waves. The RVE solution for each wave component is the same as the single-wave solution~\eqref{eq:RVEsolution} derived in the previous section. The entire solution is then obtained as the superposition of the solutions of the~$N^2$ frequencies, namely,
\begin{equation} \label{eq:RVEsolution-2}
  v^h (\vec x) 
  = 
  \frac{1}{\mu_0} 
  \sum_{k^0=0}^{N-1} \sum_{k^1=0}^{N-1} 
  \frac
  { 
    i \vec \xi_{\vec k} \cdot \hat {\vec \tau }_{\vec k} 
  }
  {
    \vec \xi_{\vec k} \cdot \vec \xi_{\vec k}
  } 
  \, e^{i \vec \xi_{\vec k} \cdot \vec x} \,  \quad \forall \, \vec \xi_{\vec k} \neq \vec 0.
\end{equation}
Finally, this approximation can be used for evaluating the fixed-point iterations~\eqref{eq:homogBVP} for the non-homogeneous problem. The resulting workflow is collected in Algorithm~\ref{alg:DFT}.

\begin{algorithm}[t] 
\caption{Fourier-based iterative solution of the RVE problem on~$\Omega = (0, \, L)^2 \subset \mathbb R^2$. \label{alg:DFT}}
{\bf Initialisation:}\\
$\vec \gamma^{(s=0)}_{\vec k} \equiv \vec \gamma^{(s=0)}_{k^0k^1} = \bar{\vec \gamma}$ \\
$\vec \sigma^{(s=0)}_{\vec k} \equiv \vec \sigma^{(s=0)}_{k^0 k^1} = \mu (\vec x_{\vec k}) \vec \gamma^{(0)}_{\vec k}$ \quad $\forall \; \vec x_{\vec k} \equiv \vec x_{k^0k^1} \in \Omega$ \\

{\bf Iteration:} $\vec \gamma^{(s)}_{\vec k}$ and $\vec \sigma^{(s)}_{\vec k}$ at every grid point~$\vec k = (k^0, \, k^1)$ are $\vec x_{\vec k}$ known \\
(a) \quad \quad \quad ${\vec \tau}^{(s)}_{\vec k} = \vec \sigma^{(s)}_{\vec k} - \mu^0 \vec \gamma^{(s)}_{\vec k} = \left (\mu(\vec x_{\vec k}) - \mu^0 \right) \vec \gamma_{\vec k}^{(s)}$ \\
(b) \quad \quad \quad $\hat {\vec \tau}_{\vec j}^{(s)} = \textrm{DFT} \left ( \vec \tau_{\vec k}^{(s)} \right ) $ \\
(c) \quad \quad \quad \textrm{Convergence \ test} \\ 
(d) \quad \quad \quad $\hat{\vec \gamma}^{{(s+1)}}_{\vec j} = \begin{cases}
     -\hat{\vec \Gamma}_{\vec j} \cdot \hat{\vec \tau}^{(s)}_{\vec j} & \text{if $j^0 \neq N/2$ and $j^1 \neq N/2$ }\\
   \sqrt{N}\bar{\vec \gamma} & \text{if $j^0 = j^1 =N/2$}\\
  \end{cases}$ \\
(e) \quad \quad \quad $\vec \gamma_{\vec k}^{(s+1)} = \textrm{DFT}^{-1} \left ( \hat{\vec \gamma}_{\vec j}^{(s+1)} \right )$\\ 
(f) \quad \quad \quad $\vec \sigma^{(s+1)}_{\vec k}  = \mu (\vec x_{\vec k}) \vec \gamma^{(s+1)}_{\vec k}$\\ 
(g) \quad \quad \quad $s \leftarrow s+1$
\end{algorithm}

%
\section{Quantum computing algorithms \label{sec:algorithms}} 
%
In this Section we present fundamental quantum algorithms and circuits that enable the formulation of quantum computing solvers for mechanics applications in general, and for the specific model application to computational homogenisation set forth in Section~\ref{sec:compHomog}. 

%
\subsection{Multi-controlled operation \label{sec:multi-controll}}
%
Conditional operations, such as the ``if-then" construct in classical computing, are essential for computational mechanics algorithms. In quantum computing, conditional operations are realised through multi-controlled gates.

The~$CNOT$ gate presented in Section \ref{sec:C-NOT} serves as a prototype for multi-controlled operations. As elaborated in that section, a controlled version of a single-qubit gate~$U$, denoted as a controlled-$U$ or $CU$ gate, is obtained by replacing the Pauli~$X$ gate in~\eqref{eq:cnot} with~$U$, i.e.,  
\begin{equation}
CU = \ket{0}\bra{0} \otimes I + \ket{1}\bra{1} \otimes U.
\end{equation}
The~$CU$ gate maps the state $\ket c \ket k$ to $\ket c U^c \ket k$. The gate~$U$ is applied to the target state~$k$ only when the control qubit is in the state~$\ket c = \ket 1$.

The generalisation of~$CU$ gates leads to multi-controlled~$C_m U$ gates with~$m$ control qubits. A~$C_m U$ gate performs the mapping
\begin{equation}
\ket{c_0c_1...c_m } \ket{k} \mapsto \ket{c_0c_1 \dotsc c_m}  U^{c_0 \cdot c_1 \cdot \dotsc \cdot c_m} \ket{k},
\end{equation}
where the gate~$U$ is applied only if all the control qubits are in the state~$\ket{c_0c_1 \dotsc c_n} = \ket{11 \dotsc 1}$. Multi-controlled gates can be implemented using, for instance, the "V-chain" construction proposed in Barenco et al.~\cite{barenco1995elementary}, see also Nielson and Chuang~\cite[Ch. 4.3]{ikeAndMike}. 

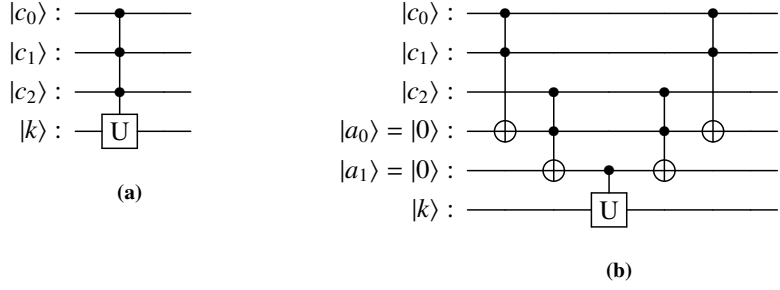
\begin{figure}
	\centering
 	\subfloat[][\label{fig:MCMT}] {	
	\scalebox{1.0}{
\Qcircuit @C=1.0em @R=0.2em @!R { \\
	 	 \lstick{\ket{{c}_{0}} : } & \ctrl{1} & \qw & \qw\\
	 	 \lstick{\ket{{c}_{1}} : } & \ctrl{1} & \qw & \qw\\
	 	 \lstick{\ket{{c}_{2}} : } & \ctrl{1} & \qw & \qw\\
	 	 \lstick{\ket{{k}} : } & \gate{\mathrm{U}} & \qw & \qw\\
\\ }}
		} \hspace{0.2\textwidth}
	\subfloat[][\label{fig:MCMT-Vchain}] {			
\scalebox{1.0}{
\Qcircuit @C=1.0em @R=0.2em @!R { \\
	 	 \lstick{\ket {{c}_{0}} : } & \ctrl{1} & \qw & \qw & \qw & \ctrl{1} & \qw & \qw\\
	 	\lstick{\ket {{c}_{1}} : } & \ctrl{2} & \qw & \qw & \qw & \ctrl{2} & \qw & \qw\\
	 	 \lstick{\ket {{c}_{2}} : } & \qw & \ctrl{1} & \qw & \ctrl{1} & \qw & \qw & \qw\\
	 	 \lstick{\ket {{a}_{0}} = \ket{0} : } & \targ & \ctrl{1} & \qw & \ctrl{1} & \targ & \qw & \qw\\
	 	 \lstick{\ket{{a}_{1}} = \ket{0} : } & \qw & \targ & \ctrl{1} & \targ & \qw & \qw & \qw\\
	 	 \lstick{\ket{{k}} : } & \qw & \qw & \gate{\mathrm{U}} & \qw & \qw & \qw & \qw\\
\\ }}
	}
\caption{Multi-controlled $C_m U$ gates. (a) A multi-controlled~$C_3 U$ gate with three control qubits $\ket{c_0c_1c_2}$ and one target qubit $\ket{k}$. (b) The V-chain implementation of the~$C_3 U$ gate using the three-qubit Toffoli gates and the~$CU$ gate. In addition, two ancillary (auxiliary) qubits~$\ket{a_0}$ and~$\ket{a_1}$ are introduced to store temporary information. }
\label{fig: MCMT_all}
\end{figure}

As an illustration, consider the~$C_3U$ gate with $m=3$ control qubits shown in Figure~\ref{fig:MCMT} and its V-chain implementation in Figure~\ref{fig:MCMT-Vchain}. The four three-qubit gates in Figure~\ref{fig:MCMT-Vchain} are Toffoli gates, which are~$CNOT$-like gates with two control qubits. The target qubit state is flipped when both control qubits of the Toffoli gate are in state~$\ket 1$. The circuit in  Figure~\ref{fig:MCMT-Vchain} contains two ancillary (auxiliary) qubits~$\ket {a_0}$ and~$\ket{a_1}$, both initialised to~$\ket 0$, to store temporary information. 
The left most Toffoli gate yields for the first ancilla~$\ket {a_0} = \ket{c_0 \cdot c_1}$, i.e.~$\ket{a_0}=\ket 1$ if and only if~$\ket{c_0} = \ket 1$ and~$\ket {c_1}= \ket 1$, else~$\ket{a_0} = \ket 0$. The second Toffoli gate yields for the second ancilla~$\ket {a_1} = \ket{a_0 \cdot c_2} = \ket{c_0\cdot c_1 \cdot c_2}$. Subsequently, a~$CU$ gate is applied when~$\ket {a_1}= \ket{c_0\cdot c_1 \cdot c_2} = \ket 1$. The last two Toffoli gates set the ancillary qubits~$\ket {a_0}$ and~$\ket {a_1}$ back to state~$\ket 0$.

\paragraph{Complexity} A general multi-controlled operation with~$m$ control qubits requires $m-2$ ancilla qubits and $2m-3$ Toffoli gates. In turn, each Toffoli gate can be decomposed into $8$ $U_3$ gates and $6$ $CNOT$ gates~\cite[Ch. 4.3]{ikeAndMike}. Recall that we use as mentioned in Section~\ref{sec:complexity} the universal gate set $\{ CNOT, \, U_3\}$.
We conclude that for a quantum system with~$n$ qubits and a state space of dimension of~$N = 2^n$, a multi-controlled gate has the linear computational complexity of $\mathcal{O} ( n)$, or $\mathcal{O} ( \log \, (N))$.  

%
 \subsection{Encoding of polynomials \label{sec:polyEncode}} 
%
The encoding of functions on a quantum computer is relevant for many computational mechanics applications. Here, we focus on the encoding of polynomials because of their approximating properties in bounded domains. 
Specifically, we consider the approach proposed in~\cite{woerner2019quantum,stamatopoulos2020option} to encode a real-valued univariate polynomial~$f(k) \in \mathbb R$ of the form
\begin{equation}
	f(k) = \sum_{j=0}^p \alpha_j k^j \, ,
\end{equation}
where~$k \in \{ 0, \, 1, \, 2, \, \dotsc, 2^{n-1} \} $ is the domain, $n$ the number of qubits, $p$ the polynomial degree and~$\alpha_j \in \mathbb R$ the prescribed coefficients. The function~$f(k)$ can be encoded by constructing a unitary~\mbox{$U_P \in \mathbb C^{2^{n+1} \times 2^{n+1}}$} that is applied to the augmented input state \mbox{$\ket k \ket 0 \equiv \ket {k_0 k_1 \dotsc k_{n-1}} \otimes \ket 0$}. This augmented state consists of the Kronecker product of the state~$\ket k$ and an ancillary (auxiliary) qubit~$\ket 0$. The ancilla allows us to embed the problem in a higher dimensional space, with the solution then determined by projecting the output state to a lower dimensional space. The unitary~$U_P$, applied to the augmented input state, yields the output state 
\begin{equation} \label{eq:polynomialRot}
	U_P \colon \ket k \ket 0 \mapsto \ket k \left (\cos( \varepsilon f(k)) \ket 0 + \sin( \varepsilon f(k)) \ket 1 \right ) = \cos (\varepsilon f(k)) \ket k \ket 0 + \sin ( \varepsilon f(k)) \ket k \ket 1  \, .
\end{equation}
Thus, $U_P$ makes use of the rotation gate~$R_Y(2 \varepsilon f(k))$ and concerns only the ancillary qubit originally in state~$\ket 0$. Intuitively,~$U_P$ applies a rotation by~$\varepsilon f(k)$ in the plane spanned by the right-most qubit. The parameter $\varepsilon \in \mathbb R$ is a small scaling factor, which is chosen such that~\mbox{$\varepsilon f(k) \ll 1$} so that the linearisation of the sine term in~\eqref{eq:polynomialRotApprox} gives the approximation
\begin{equation} \label{eq:polynomialRotApprox}
	\cos (\varepsilon f(k)) \ket k \ket 0 + \sin ( \varepsilon f(k)) \ket k \ket 1 \approx  \sqrt{1-(\varepsilon f(k))^2} \ket k \ket 0 + \varepsilon f(k) \ket k \ket 1  \, .
\end{equation}
That is, for sufficiently small~$\varepsilon $, the amplitude is equal to the sought function value~$\varepsilon f(k) $ when the ancillary qubit is in state~$\ket 1$. In contrast, when the ancillary qubit is in state~$\ket 0$ the amplitude is of no interest and can be discarded. As already mentioned, the rotation gate~$R_Y(2 \varepsilon f(k))$ is applied only to the ancillary qubit originally in state~$\ket 0$. Because~$f(k)$ is additively composed of~$p$ scaled monomials, the gate~$R_Y(2 \varepsilon f(k))$ can be multiplicatively composed of the~$p$ rotations~$R_Y(2 \varepsilon  a_m k^m)$. Furthermore, the order of the rotations~$R_Y(2 \varepsilon  \alpha_m k^m)$ is irrelevant because all the rotations pertain to the same plane spanned by the ancillary qubit.

\begin{figure}
  \centering
\scalebox{0.975}{
\Qcircuit @C=1.em @R=0.2em @!R { 
	 	\nghost{\,} &\lstick{ \ket{k_0} : } & \ctrl{3} & \qw & \qw & \ctrl{3} & \qw & \ctrl{3} & \qw & \qw\\
	 	\nghost{\, }& \lstick{\ket{k_1} : } & \qw & \ctrl{2} & \qw & \qw & \ctrl{2} & \ctrl{2} & \qw & \qw\\
	 	\nghost{\,}& \lstick{\ket{k_2} : } & \qw & \qw & \ctrl{1} & \ctrl{1} & \ctrl{1} & \qw & \qw & \qw\\
       \nghost{\, }& \lstick{\ket 0 : } & \gate{R_Y( 8(\alpha_1 + 4 \alpha_2) )} & \gate{R_Y( 4(\alpha_1 + 2 \alpha_2) )} & \gate{R_Y( 2(\alpha_1 + \alpha_2) )} & \gate{R_Y( 16 \alpha_2)} & \gate{R_Y( 8 \alpha_2)}  & \gate{R_Y( 32 \alpha_2)} & \gate{R_Y( 2 \alpha_0 )} & \qw\\
 }}
\caption{Quantum circuit for implementing the mapping~\mbox{$\ket k \ket 0 \mapsto \cos (f(k)) \ket k \ket 0 + \sin ( f(k)) \ket k \ket 1 $}. Here,~\mbox{$\ket k= \ket{k_0 k_1 k_2}$} is represented with three qubits and the ancillary qubit is initially in state~$\ket 0$. For instance, providing the circuit with the input~\mbox{$\ket{k} = \ket 1 \ket 1 \ket 0$}, i.~e., \mbox{$k=1 \cdot 2^2+ 1 \cdot 2^1 + 0 \cdot 2^0= 6$}, it will yield the output~\mbox{$\cos (f(6)) \ket 1 \ket 1 \ket 0 \ket 0 + \sin ( f(6)) \ket 1 \ket 1 \ket 0 \ket 1$}.  \label{fig:polyEncode}} 
\end{figure}
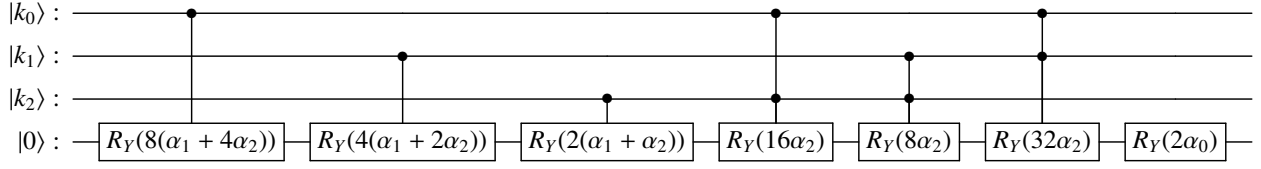

The implementation of this sketched algorithm as a quantum circuit can be illustrated as follows. Consider, by way of example, the encoding of the quadratic polynomial 
\begin{equation} \label{eq:function_quad}
	f(k) = \alpha_2 k^2 + \alpha_1 k + \alpha_0 \, .
\end{equation}
The number of qubits~$n$ for representing~$k$ depends on the application. We choose here~$n=3$ so that~$k$ has the binary representation 
\begin{equation}\label{eq:birep}
	k = k_0k_1k_2 = k_0 2^2 + k_1 2^1 + k_2 2^0 \, .
\end{equation}
We note that $k_0^2 = k_0$, $k_1^2 = k_1$ and $k_2^2 = k_2$, since each bit is either~$0$ or~$1$. Introducing (\ref{eq:birep}) into~\eqref{eq:function_quad} gives the identity
\begin{equation}
	f(k) = 
   4(\alpha_1 + 4 \alpha_2) k_0 + 2( \alpha_1 + 2 \alpha_2 ) k_1 + ( \alpha_1 + \alpha_2 ) k_2 +  8 \alpha_2 k_0 k_2 + 4 \alpha_2 k_1 k_2 + 16 \alpha_2 k_0 k_1  + \alpha_0 \, .
\end{equation}
To begin the encoding process, we start with a 4-qubit system $|k_0k_1k_2\rangle|0\rangle$. Recalling the definition of the $R_Y$ gate~\eqref{eq:rotationGates}, for arbitrary $\theta \in \mathbb{R}$, we have:
\begin{equation}
  R_Y(2\theta)|0\rangle 
  = 
  e^{-i\theta Y} |0\rangle 
  = 
  \cos(\theta)|0\rangle + \sin(\theta)|1\rangle \, . 
\end{equation}
Let $\theta = f(k)$, where we omit the constant $\varepsilon $ for simplicity. The goal is to implement the operator $e^{-if(k)Y}$. Conveniently, the sum in the polynomial $f(k)$ is converted into a product by the exponential map, namely, 
\begin{align} \label{eq:Ry_exponential}
\begin{split}
  e^{-if(k)Y}
  &= 
  e^{-4 i (\alpha_1+4\alpha_2)k_0 Y} \,
  e^{-2 i (\alpha_1+2\alpha_2)k_1 Y} \,
  e^{-i(\alpha_1+\alpha_2)k_2 Y} \,
  e^{-8 i\alpha_2k_0k_2 Y} \,
  e^{-4 i \alpha_2k_1k_2 Y} \,
  e^{-16 i \alpha_2k_0k_1 Y} \,
  e^{-i\alpha_0Y} \\
  & = 
  R_Y(8(\alpha_1+4\alpha_2)k_0) \,
  R_Y(4(\alpha_1+2a_2)k_1) \,
  R_Y(2(\alpha_1+\alpha_2)k_2) \, 
  R_Y(16\alpha_2k_0k_2) \,
  R_Y(8\alpha_2k_1k_2) \,  \\
  & \times
  R_Y(32\alpha_2k_0k_1) \,
  R_Y(2\alpha_0) .
\end{split}
\end{align}
We can thus implement the operator using multi-controlled $R_Y$ gates. For example, we may apply~$R_Y(8 (\alpha_1+ 4 \alpha_2)$ to the ancillary qubit only if~$\ket{k_0} = \ket{1}$ or~$R_Y(4 (\alpha_1+ 2 \alpha_2))$ only if $\ket{k_1} = \ket{1}$. In this manner, we can encode the full polynomial by successively applying multi-controlled $R_Y$ gates as shown in the circuit diagram in Figure \ref{fig:polyEncode}.

The preceding technique can be easily generalised to multivariate polynomials. Consider, for instance, the bivariate polynomial
\begin{equation}~\label{eq:2D Polynomial}
	f(\vec k)  = \sum_{\vec m = \vec 0}^p \alpha_{\vec m} \vec k^{\vec m}  = \sum_{m^0=0}^{p}  \sum_{m^1=0}^{p}  \alpha_{m^0 m^1} (k^0)^{m^0} (k^1)^{m^1} \, ,
\end{equation}
where~$\vec k=(k^0, \, k^1)$ and~$\vec m=(m^0, \, m^1)$ are multi-indices,~$p$ the is polynomial degree, and~$\alpha_{\vec m} \in \mathbb R$ are the coefficients. The argument~$\vec k$ is represented with~$n^2$ qubits, i.e.,~\mbox{$k^0, k^1 \in \{ 0, \, 1,  \, 2, \dotsc , \,  2^{n-1}\}$.}  The augmented unitary~$U_P \in \mathbb C^{2 ^{n^2+1} \times 2^{n^2+1}}$ now implements the mapping 
\begin{equation}~\label{eq:polynomialRot2D}
	U_P \colon   \ket{\vec k} \ket 0   \mapsto \cos (\varepsilon f(\vec k)) \ket{\vec k} \ket 0 + \sin (\varepsilon f (\vec k)  ) \ket {\vec k} \ket 1 \, .
\end{equation}
We express the argument~$\vec k$ as a binary~$\vec k = ( k^0_0 k^0_1\dotsc k^0_{n-1}, \,  k^1_0 k^1_1\dotsc k^1_{n-1}) $. Substituting it into~\eqref{eq:2D Polynomial} and making use of the multinomial expansion theorem yields
\begin{align} 
\begin{split}
f(\vec k) &= \sum_{\vec m =\vec 0}^p \alpha_{\vec m} \left (k_0^0 2^{n-1}+k_1^0 2^{n-2}+ \dotsc + k^0_{n-1} 2^0  \right )^{m_0}  \left ( k_0^1 2^{n-1}+ k^1_1  2^{n-2}  +\dotsc +  k^1_{n-1} 2^0 \right)^{m_1}\\
& = \sum_{\vec m= \vec 0}^p\alpha_{\vec m}   \left (  \sum_{j_0^0+j^0_1+..+j^0_{n-1}=m^0}  \frac{m^0!}{j^0_0! j^0_1! \dotsc j^0_{n-1}!} \prod_{s=0}^{n-1} \left (2^{n-1-s} k^0_{s} \right )^{j^0_s}   \right ) 
\\
 &\phantom{  = \sum_{\vec m=0}^p   \; \; \, }  \times \left (  \sum_{j_0^1+j^1_1+...+j^1_{n-1}=m^1} \frac{m^1!}{j^1_0! j^1_1! \dotsc j^1_{n-1}!}   \prod_{s=0}^{n-1} \left (2^{n-1-s} k^1_{s} \right )^{j^1_s}  \right )  \, .
\end{split} \, 
\end{align}
The corresponding quantum circuit consists of multi-controlled~$R_Y$ gates conditioned on up to~$n^2$ qubits, with rotation angles twice of
\begin{equation}
	\alpha_{m^0m^1}  \frac{m^0!}{j^0_0! j^0_1! \dotsc j^0_{n-1}!}  \frac{m^1!}{j^1_0! j^1_1! \dotsc j^1_{n-1}!}   \prod_{s=0}^{n-1} \left (2^{n-1-s} \right )^{j^0_s}  \prod_{s=0}^{n-1} \left (2^{n-1-s}  \right )^{j^1_s}  \, .
\end{equation}

\paragraph{Complexity}  For a univariate polynomial, we note that the number of terms in $(2^{n-1}k_{0}+ 2^{n-2}k_{1} + \dotsc + 2^0k_{n-1})^m$ for a fixed $m$ is equal to 
\begin{equation}
	\binom{n+m-1}{n-1}
\end{equation}
Hence, the total number of terms for a polynomial of degree~$p$ with $m \in \{0,1,..., p\}$ is 
\begin{equation}  \label{eq: sum multinomial}
  \binom{n-1}{n-1}+\binom{n}{n-1}+\binom{n+1}{n-1}+...+\binom{n+p-1}{n-1}   = \binom{n+p}{n} \, .
\end{equation}
Thus, we conclude that for the encoding of uni-  and bivariate functions a total of $\binom{n+p}{n}$ and $\binom{n+p}{n}^2$ multi-controlled $R_Y$ gates are required. Consequently, polynomial encoding can be implemented with $\mathcal{O} \left (\poly (pn) \right )$ for univariate polynomials and $\mathcal{O} \left (\poly (pqn^2) \right )$ for bivariate polynomials. 
 
%
 \subsection{State preparation via function encoding \label{sec:statePrepEF}}
%
On a quantum computer, a classical vector~$\vec q \in \mathbb C^N$ with complex Euclidean norm~$\| \vec q \| = 1$ can be encoded as a quantum state $\ket q$ using $n=\log_2 N$ qubits. This means, for instance, that ~$n=3$ qubits are sufficient to encode a classical vector with~$N=8$ components. Considering that each of the~$n=3$ qubits has only two states, the encoding of a vector with~$N=8$ components requires entanglement. The encoding of a classical vector into a quantum state is referred to as state preparation. 

There are a number of approaches and algorithms for efficient state preparation~\cite{mottonen2004transformation,shende2005synthesis,araujo2021divide}. In all algorithms a quantum state~$\ket q$ is prepared by applying a suitably constructed unitary~$U_S$ to the zero-state, i.~e., 
\begin{equation}
	U_S \colon \ket 0^{\otimes n } = \ket {\underbrace{0 \dotsc 0}_n} \mapsto \ket q = \sum_{k=0}^{2^n-1} q_k\ket k \, , 
\end{equation}
so that the amplitudes~$q_k$ are equal to the components of the classical vector we seek to encode. 

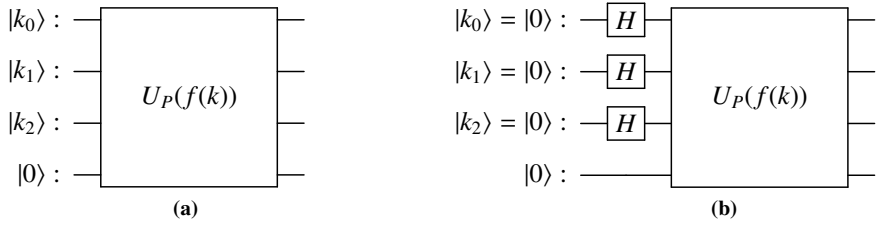
\begin{figure}
	\centering
 	\subfloat[][\label{fig:polyStatePrepA}] {	
\scalebox{1.0}{
\Qcircuit @C=1.0em @R=1.05em @!R { \\
	 	 \lstick{\ket {k_0} : } & \multigate{3}{\quad U_P(f(k)) \quad } & \qw \\
	 	 \lstick{\ket{k_1} : } & \ghost{\quad U_P(f(k)) \quad} & \qw \\
	 	 \lstick{\ket{k_2} : } & \ghost{\quad U_P(f(k)) \quad} & \qw \\
	 	 \lstick{\ket{0} : } & \ghost{\quad U_P(f(k)) \quad} & \qw \\
 }}
		} \hspace{0.2\textwidth}
	\subfloat[][\label{fig:polyStatePrepB}] {			
\scalebox{1.0}{
\Qcircuit @C=1.0em @R=0.7em @!R { \\
	 	\lstick{\ket {k_0} = \ket 0 : } & \gate{H} & \multigate{3}{ \quad U_P(f(k)) \quad } & \qw \\
	 	\lstick{\ket{k_1} =\ket 0 : } & \gate{H} & \ghost{\quad U_P(f(k)) \quad } & \qw \\
	 	 \lstick{\ket{k_2} = \ket 0 : } & \gate{H} & \ghost{\quad U_P(f(k)) \quad } & \qw \\
	 	 \lstick{\ket{0} : } & \qw & \ghost{\quad U_P(f(k)) \quad } & \qw \\
 }}
	}
\caption{Quantum state preparation via function encoding. The four-qubit gate in (a) represents the unitary~$U_P(f(k))$ for encoding~$f(k)$ and is composed as the circuit depicted in Figure~\ref{fig:polyEncode}. Prepending the circuit with three Hadamard gates as shown in (b) yields a uniform superposition state, which becomes after multiplication by~$U_P(f(k))$ an entangled state. \label{fig:polyStatePrep}}\end{figure}

It is possible to use the polynomial encoding approach introduced in the previous section to devise a technique for constructing an approximate unitary. To this end, we first determine a polynomial such that~$f(k) \approx q_k$, for all~$k \in \{ 0, \, 1,\, \dotsc , \, N-1 \} $, approximating the components of a classical vector~$\ket q \in \mathbb R^N$, on a classical computer using standard polynomial approximation techniques, specifically Chebyshev interpolation, see also Section~\ref{sec:encode_function}. After~$f(k)$ is determined, we use the method described in the previous section to construct a unitary~$U_P$ and to evaluate it for different~$k$ on a quantum computer, cf.~\eqref{eq:polynomialRot}. Crucially, quantum superposition enables the evaluation of~$f(k)$ for all possible values of~$k$ simultaneously. To this end, before applying the unitary~$U_P$ we first generate a uniform quantum state composed of all basis vectors. As illustrated in Figure~\ref{fig:polyStatePrepB} for~$n=3$ a uniform quantum state is obtained by applying Hadamard gates. More specifically, the three Hadamard gates in Figure~\ref{fig:polyStatePrepB} accomplish the following transformation:
\begin{align}
\begin{split}
	(H \otimes H \otimes H \otimes I ) \colon  \ket 0 \otimes \ket 0 \otimes \ket 0 \otimes \ket 0 & \mapsto \frac{1}{2 \sqrt 2} \left ( \ket 0 + \ket 1 \right ) \otimes \left ( \ket 0 + \ket 1 \right ) \otimes \left ( \ket 0 + \ket 1 \right ) \otimes \ket 0  \\
	 & = \frac{1}{2 \sqrt 2} \sum_{k_0=0}^1  \sum_{k_1=0}^1  \sum_{k_2=0}^1 \ket {k_0} \ket{k_1} \ket{k_2} \ket 0 = \frac{1}{2 \sqrt 2}  \sum_{k=0}^7 \ket k \ket 0 \, . 
\end{split}
\end{align}
According to~\eqref{eq:polynomialRot} and~\eqref{eq:polynomialRotApprox}, the application of the unitary~$U_P \left (\varepsilon f(k) \right )$ to this state yields 
\begin{align} \label{eq:polyStatePrep}
	\begin{split}
	U_P \left (\varepsilon f(k) \right ) \colon \frac{1}{2 \sqrt 2} \sum_{k=0}^7 \ket k \ket 0 & \mapsto \frac{1}{2 \sqrt 2} \sum_{k=0}^7 \left ( \cos (\varepsilon f(k)) \ket k \ket 0 + \sin ( \varepsilon f(k)) \ket k \ket 1 \right ) \\ 
		& \approx \frac{1}{2 \sqrt 2} \sum_{k=0}^7 \left (\sqrt{1-(\varepsilon f(k))^2} \ket k \ket 0 + \varepsilon f(k) \ket k \ket 1 \right ) \, ,
	\end{split}
\end{align}
as desired. 

%
\subsection{An exact state preparation algorithm} \label{sec:state initial}
%
An alternative and exact state preparation approach motivated by M\"ott\"onen et al.~\cite{mottonen2004transformation}, see also~\cite{araujo2021divide}, and closer to algorithms available in quantum computing SDKs, such as Qiskit, is the following. To illustrate the basic idea, we consider the encoding of a classical vector~$\vec q \in \mathbb R^8$, $N=8$, using the amplitudes of~$n=3$ qubits. The quantum state is prepared using the unitary mapping 
\begin{equation}
	 U_S \colon \ket {000}  \mapsto \ket q  \, . 
\end{equation}
However, it proves convenient to first compute the inverse mapping 
\begin{equation} \label{eq:statePrepInv}
	 U_S^\dagger \colon \ket {q} \mapsto  \ket{000} \, . 
\end{equation}
and then compute $U_S$ simply by taking its conjugate transpose. A suitable unitary~$U_S^\dagger $ can be constructed by applying either successive Givens rotations or Householder reflections, which are well-known from linear algebra, see e.g.~\cite{strangLinAlg}. We briefly sketch how~$U_S^\dagger $ can be constructed using Givens rotations. The mapping~\eqref{eq:statePrepInv}, expressed in matrix notation, takes the form
\begin{equation} \label{eq:givensMatrix1}
    \begin{pmatrix} \cdot & \cdot & \cdot & \cdot & \cdot & \cdot & \cdot & \cdot \\ \cdot & \cdot & \cdot & \cdot & \cdot & \cdot & \cdot & \cdot \\ \cdot & \cdot & \cdot & \cdot & \cdot & \cdot & \cdot & \cdot \\ \cdot & \cdot & \cdot & \cdot & \cdot & \cdot & \cdot & \cdot \\ \cdot & \cdot & \cdot & \cdot & \cdot & \cdot & \cdot & \cdot \\ \cdot & \cdot & \cdot & \cdot & \cdot & \cdot & \cdot & \cdot \\ \cdot & \cdot & \cdot & \cdot & \cdot & \cdot & \cdot & \cdot \\ \cdot & \cdot & \cdot & \cdot & \cdot & \cdot & \cdot & \cdot \end{pmatrix} \begin{pmatrix} q_0 \\ q_1 \\ q_2 \\ q_3 \\ q_4 \\ q_5 \\ q_6 \\ q_7 \end{pmatrix} = 
   \begin{pmatrix} 1 \\ 0 \\ 0 \\ 0 \\ 0 \\ 0 \\ 0 \\ 0 \end{pmatrix}\, , 
\end{equation}
where the yet unknown components are represented by a symbol~"$\cdot$". Each Givens rotation represents a rotation in the plane spanned by two coordinate axes. In the first step, the rotation axis are chosen such that 
\begin{equation}
 	 \begin{pmatrix*}[r] c & s & 0& 0 & 0 & 0 & 0 & 0 \\ -s & c & 0 & 0 & 0 & 0 & 0 &0 \\ 0 & 0 & \cdot & \cdot &0 & 0 &0 & 0 \\ 0 & 0 & \cdot & \cdot & 0 & 0 & 0 &0 \\ 0 & 0 & 0 & 0 & \cdot & \cdot & 0 & 0 \\ 0 & 0 & 0 & 0 & \cdot & \cdot & 0 & 0 \\ 0 & 0 & 0 & 0 & 0 & 0 & \cdot & \cdot \\ 0 & 0 & 0 & 0 & 0 & 0 & \cdot & \cdot \end{pmatrix*} \begin{pmatrix} q_0 \\ q_1 \\ q_2 \\ q_3 \\ q_4 \\ q_5 \\ q_6 \\ q_7 \end{pmatrix} = \begin{pmatrix} r \\ 0 \\ \cdot \\ 0 \\  \cdot \\ 0 \\ \cdot \\ 0 \end{pmatrix} \, .
\end{equation}
Some of the non-zero components have been omitted for clarity. Note that on the right-hand side every other component is zero. The non-zero components of~$U_s^\dagger$ in the upper left~$2\times2$ block are 
\begin{equation}
	r = \sqrt{q_0^2 + q_1^2 }	\, , \quad  c = \cos \beta = \frac{q_0}{r}  \, , \quad s = \sin \beta = \frac{q_1}{r}  \, .
\end{equation}
The components of the other three~$2 \times 2$ blocks are determined similarly. In the second step, the Givens rotations applied are chosen such that
\begin{equation} \label{eq:givensMatrix2}
	 \begin{pmatrix*}[r] \tilde c & 0 &  \tilde s& 0 & 0 & 0 & 0 & 0 \\ 0 & 0 & 0 & 0 & 0 & 0 & 0 &0 \\ - \tilde s & 0 & \tilde c & 0 &0 & 0 &0 & 0 \\ 0 & 0 & 0 & 0 & 0 & 0 & 0 &0 \\ 0 & 0 & 0 & 0 & \cdot & 0 & \cdot & 0 \\ 0 & 0 & 0 & 0 & 0 & 0 & 0 & 0 \\ 0 & 0 & 0 & 0 & \cdot & 0 & \cdot & 0 \\ 0 & 0 & 0 & 0 & 0 & 0 & 0 & 0 \end{pmatrix*} \begin{pmatrix}  r \\ 0 \\ \cdot \\ 0 \\  \cdot \\ 0 \\ \cdot \\ 0 \end{pmatrix} = \begin{pmatrix} \tilde r \\ 0 \\ 0 \\ 0 \\ \cdot \\ 0 \\ 0 \\ 0 \end{pmatrix}  \, .
\end{equation}
Notice that the right-hand side now has only two non-zero components. In the third and last step, the fifth component of the right-hand side can be set zero by applying one more Givens rotation yielding a vector with a~$1$ on its first component, i.e.~$\ket{000}$. 

\begin{figure}[]
\centering
\scalebox{1.0}{
\Qcircuit @C=1.0em @R=0.2em @!R { \\
	 	 \lstick{\ket{k_0} : } & \qw & \ctrlo{1} &  \ctrlo{1}&  \ctrl{1} &  \ctrl{1} \barrier[0em]{2} & \qw & \ctrlo{1} & \ctrl{1} \barrier[0em]{2} & \qw & \gate{R_Y(\mathrm{\cdot})} & \qw & \qw\\
	 	\lstick{\ket{k_1} : }  & \qw & \ctrlo{1} & \ctrl{1} & \ctrlo{1} & \ctrl{1} & \qw & \gate{R_Y\mathrm{(-2 \tilde \beta})} & \gate{R_Y(\mathrm{\cdot})} & \qw & \ctrlo{-1} & \qw & \qw\\
	 	\lstick{\ket{k_2} : } & \qw & \gate{R_Y(\mathrm{-2 \beta })} & \gate{R_Y(\mathrm{\cdot})}  &\gate{R_Y(\mathrm{\cdot})} & \gate{R_Y(\mathrm{\cdot})} & \qw & \ctrlo{-1} & \ctrlo{-1} & \qw & \ctrlo{-1} & \qw & \qw\\
 }}
\caption{Quantum circuit implementing the unitary~$U_S^\dagger$ for mapping an arbitrary (real) state~$\ket q = ( q_0 \; q_1 \; q_2 \; q_3 \; q_4 \; q_5 \; q_6 \; q_7 )^\trans$ to a state~\mbox{$\ket{000} = ( 1 \; 0 \; 0 \; 0 \; 0 \; 0 \; 0 \; 0)^\trans $}. The two dashed helper lines correspond to the three distinct steps in constructing~$U_S^\dagger$ mentioned in the main text. Notice that each~$R_Y$ is applied only to two selected components of the state vector. For instance, the left most~$R_Y$ gate is applied to the components~$q_0$ and~$q_1$, i.e. when~$\ket{k_0}=\ket{k_1} = \ket 0$, the second left most~$R_Y$ gate to the components~$q_2$ and~$q_3$, i.e. when~$\ket{k_0}=\ket 0$ and~$\ket {k_1} = \ket 1$ , etc. 
The inverse circuit~$U_S$ for state preparation maps the state~$\ket {000}$ to~$\ket q$. The circuit for~$U_S$ is obtained by taking the conjugate transpose~$R_Y^\dagger$ of each gate and applying the gates from right to the left. \label{fig:statePrep}}
\end{figure}
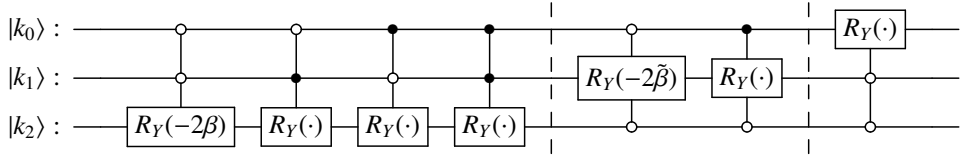

The unitary matrix~$U_S^\dagger$ obtained is the product of three unitary matrices, each containing one or more Givens rotations. Specifically, the first matrix~\eqref{eq:givensMatrix1} contains three, the second matrix~\eqref{eq:givensMatrix2} two and the third matrix one Givens rotation. On a quantum computer every Givens rotation is realised using the~$R_Y$ gate~\eqref{eq:rotationGates}. In order to apply each rotation to selected components of a vector we use the controlled versions of the~$R_Y$ gates. That is, the~$R_Y$ gate is only applied when the control qubit is in state~$\ket 1$. The circuit implementation of the construction of~$U_S^\dagger$ is depicted in Figure~\ref{fig:statePrep}. The circuit diagram for computing the inverse mapping~$U_S$ is obtained by taking the conjugate transpose of each~$R_Y$ and applying the gates from right to the left.

%
 \subsection{Encoding of arbitrary functions} \label{sec:encode_function}
%
The accurate approximation of arbitrary functions by polynomials is an extensively studied topic in numerical analysis and a number of efficient algorithms and implementations are available~\cite{berrut2004barycentric}. An arbitrary function~\mbox{$r(k) \in \mathbb R$} can be encoded on a quantum computer by first determining its piecewise polynomial approximation and then encoding the obtained polynomials. We perform the piecewise polynomial approximation on a classical computer using the Python Numpy library. Subsequently, we encode the resulting polynomials using the algorithm described in Section~\ref{sec:polyEncode}. 

In piecewise approximation, the polynomial degree~$p$ in the intervals and the number of the subintervals~$n_\omega$ are both fixed and chosen empirically. The polynomials of degree~$p$ are determined so that they interpolate the given function~$r(k)$ at~$p+1$ collocation points. As is well-known, choosing Chebyshev nodes as the collocation points yields a particularly efficient approximation with small interpolation errors. Furthermore, at the Chebyshev nodes of the first kind the Chebyshev polynomials satisfy a certain orthogonality relationship so that the interpolation can be performed without inverting a Vandermonde matrix. 

After determining the polynomial in the Chebyshev basis it is straightforward to convert it into a monomial basis as required for encoding. Furthermore, if~$n_\omega >1$, it is necessary to determine in which interval the argument~$k$ of $r(k)$ lies in order to evaluate the polynomial approximant corresponding to the interval. The interval number is determined by comparing the position of~$k$ with respect to the interval boundaries. 

\begin{figure}
	\centering
	\subfloat[][\label{fig:oneOverRA}]{
	\includegraphics[width=0.4\textwidth]{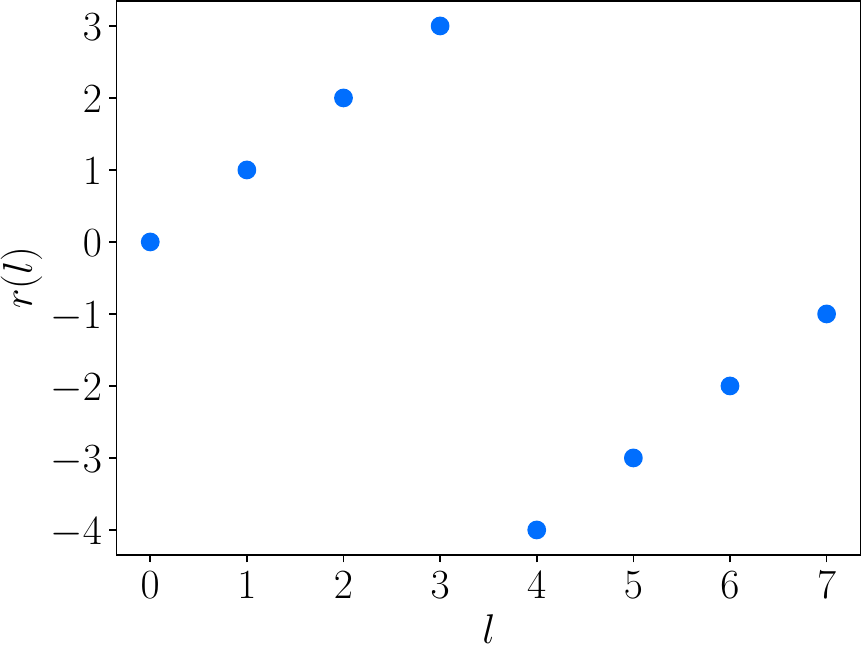}} \hspace{0.1\textwidth}
	\subfloat[][\label{fig:oneOverRB}]{
	\includegraphics[width=0.4\textwidth]{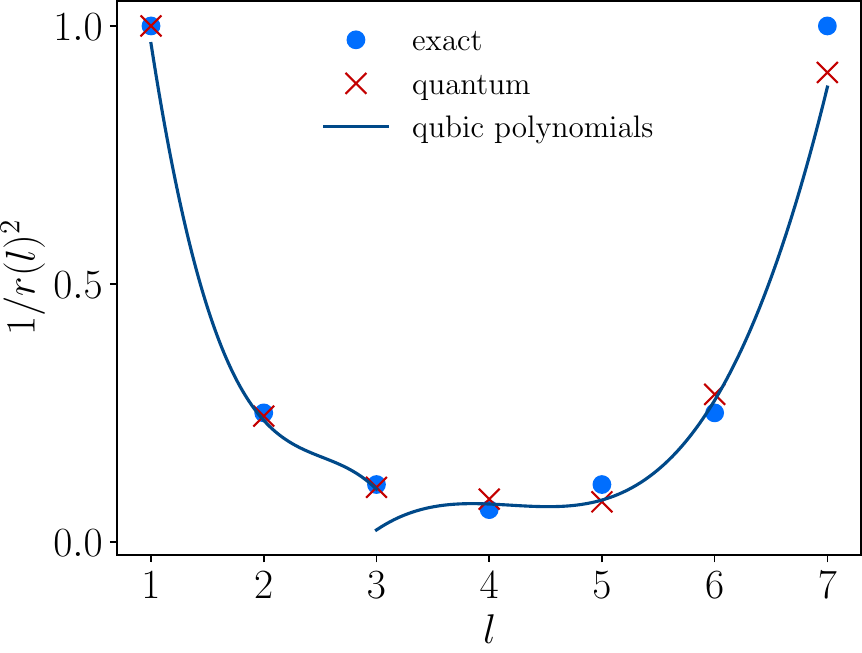}}
	\caption{Mappings of the~$N=8$ components of the state vector. In (a) the mapping of the components to wave numbers~$r(k)$ according to~\eqref{eq:componentsToFrequencies} is plotted. In (b) the function~$1/r(k)^2$ needed for computing the second derivatives and its approximation are illustrated. The dots ($\bullet$) denote the exact values which are approximated with two qubic polynomials in the two intervals $[1, \, 3)$ and $[3, \, 7]$. The polynomials are obtained by interpolating~$1/r(k)^2$ at the four Chebyshev points of the first kind. The four Chebyshev points in the two intervals are not shown. The crosses ($\times$) denote the values obtained from the quantum circuit in Figure~\ref{fig:circEvalPiecewise}. \label{fig:oneOverR2}}
\end{figure}

We further illustrate the approach by means of an example. In homogenisation the inverse square of the function~$r(k)$ introduced in~\eqref{eq:componentsToFrequencies} is required, which is discontinuous at the centre of the interval~$[0, \, 7]$, Figure~\ref{fig:oneOverRA}. Note that~$r(k)$ corresponds to a physical domain discretised with~$N=8$ grid cells and the respective nodal vectors have~$N=8$ components, so that they require~$n=\log_2 8=3$ qubits to represent on a quantum computer. The square inverse~$1/(r(k))^2$ is plotted in Figure~\ref{fig:oneOverRB}. We approximate the square inverse by means of two polynomials~$f^0(k)$ and~$f^0(k)$ of degree~$p=3$ over the intervals~$\omega^1=[1, \,3]$ and $\omega^2=[3, \, 7]$, i.e.~$n_\omega = 2$. Evidently, the accuracy of the approximation can be improved as much as desired by increasing the polynomial degree~$p$ and the number of subintervals~$n_\omega$.   

\begin{figure}
  \centering
\scalebox{0.975}{
\Qcircuit @C=1.0em @R=0.2em @!R { \\
	 	 \lstick{\ket {k_0} : } & \multigate{6}{U_{C_0}} & \multigate{3}{U_{P_0}} & \multigate{6}{U_{C_0}^\dagger} & \multigate{6}{U_{C_1}} & \multigate{3}{U_{P_1}} & \multigate{6}{U_{C_1}^\dagger} & \qw \\
	 	 \lstick{ \ket{k_1}  : } & \ghost{U_{C_0}} & \ghost{U_{P_0}} & \ghost{U_{C_0}^\dagger} & \ghost{U_{C_1}} & \ghost{U_{P_1}} & \ghost{U_{C_1}^\dagger}   &\qw \\
	 	 \lstick{ \ket {k_2}  : } & \ghost{U_{C_0}} & \ghost{U_{P_0}} & \ghost{U_{C_0}^\dagger} & \ghost{U_{C_1}} & \ghost{U_{P_1}} & \ghost{U_{C_1}^\dagger}     & \qw \\
	 	\lstick{\ket {a_0} = \ket 0 : } & \ghost{U_{C_0}} & \ghost{U_{P_0}} & \ghost{U_{C_0}^\dagger}& \ghost{U_{C_1}} & \ghost{U_{P_1}} & \ghost{U_{C_1}^\dagger}   &  \qw \\ 
	 	 \lstick{\ket {a_1}= \ket 0: } & \ghost{U_{C_0}} &\ctrl{-1} & \ghost{U_{C_0}^\dagger} & \ghost{U_{C_1}} &\ctrl{-1} & \ghost{U_{C_1}^\dagger}  & \qw \\
	 	\lstick{ \ket 0 : } & \ghost{U_{C_0}} & \qw & \ghost{U_{C_0}^\dagger} & \ghost{U_{C_1}} & \qw & \ghost{U_{C_1}^\dagger}   & \qw \\
	 	 \lstick{ \ket 0 : } & \ghost{U_{C_0}} &\qw & \ghost{U_{C_0}^\dagger} & \ghost{U_{C_1}} & \qw & \ghost{U_{C_1}^\dagger} & \qw\\
\\ }} 
\caption{Quantum circuit for implementing the mapping~\mbox{$\ket {k} \ket {0}^{\otimes 4} \mapsto \cos (r(k)) \ket k \ket 0 \ket 0^{\otimes 3} + \sin ( r(k)) \ket k \ket 1 \ket 0^{\otimes 3} $}. This circuit is constructed using the qiskit PiecewiseChebyshev class and encodes the function~$r(k)$ plotted in Figure~\ref{fig:oneOverRB}. The function~$r(k)$ is defined over two intervals. The two unitaries~$U_{C_0}$ and~$U_{C_1}$ set the the ancillary qubit~$\ket{a_1}$ to state~$\ket 1$ when the evaluation point~$\ket{k}$ lies in their respective intervals. The corresponding unitaries $U_{C_0}^\dagger$ and~$U_{C_1}^\dagger$ set the ancillary qubit~$\ket{a_1}$ back to state~$\ket 0$. The polynomials for the two intervals are evaluated with the controlled unitaries~$U_{P_0}$ and~$U_{P_1}$ conditioned on the ancillary qubit~$\ket{a_1}$. After the circuit is executed the function value~$\sin (r(k)) \ket k$ is encoded in the top three qubits when the ancillary qubit~$\ket{a_0}$ is found in state~$\ket{1}$. The remaining two ancillary qubits are only used temporarily by the circuit during execution and are of no relevance. \label{fig:circEvalPiecewise} } 
\end{figure}
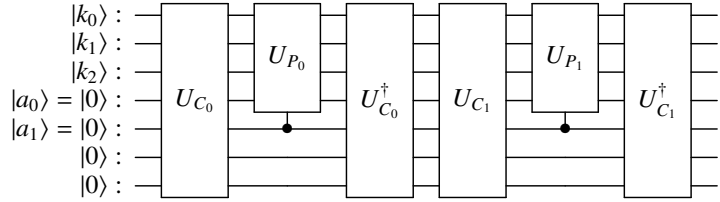

The quantum circuit for evaluating a piecewise polynomial approximation is shown in Figure~\ref{fig:circEvalPiecewise}. The top three qubits~$\ket{k_0}$, $\ket{k_1}$ and~$\ket{k_2}$ represent the state vector and the remaining four qubits are ancillary qubits. After execution, the function value $r(k)$ is equal to the amplitude of the second component of the ancillary qubit~$\ket{a_0}$. The two gates~$U_{P_0}$ and~$U_{P_1}$ represent the circuits for evaluating the two polynomials~$f^0(k)$ and~$f^1(k)$ and are constructed as described in Section~\ref{sec:polyEncode}. Note that~$U_{P_0}$ and~$U_{P_1}$ are conditioned on the ancilla~$\ket{a_1}$ and are executed only when~$\ket{a_1} = \ket{1}$. The two gates~$U_{C_0}$ and~$U_{C_1}$ determine whether~$k$ is in interval~$\omega^1$ or~$\omega^2$ and set~$\ket{a_1}$ accordingly to the state~$\ket{1}$. This is achieved by integer comparison of~$k$ and the interval boundaries. The inverse gates~$U_{C_0}^\dagger$ and~$U_{C_1}^\dagger$ reset, or unset, the~$\ket{a_1}$ and the bottom two ancilla qubits to their original state~$\ket 0$. The construction of the subcircuits~$U_{C_0}$ and~$U_{C_1}$ is outlined in~\cite{haner2018optimizing}. Their inverses are obtained by simply executing the subcircuits backwards. See also~\cite{vazquez2022enhancing} for the overall circuit for encoding arbitrary functions. The sketched circuit is available in the Qiskit PiecewiseChebyshev class. 

\paragraph{Complexity} We observe that the operators $U_{C_0}$ and $U_{C_1}$ can be effectively implemented using quantum bit string comparator algorithms, as elaborated in Oliveira et al.~\cite{oliveira2007quantum}. This reference shows that a subtraction-based algorithm facilitates the construction of these operators, with the complexity scaling according to $\mathcal{O}(\textrm{Poly}(n))$. Furthermore, as discussed in Section \ref{sec:polyEncode}, we implement the operators $U_{P_0}$ and $U_{P_1}$ using the polynomial encoding algorithm. The controlled versions of $U_{P_0}$ and $U_{P_1}$ are realised through the V-chain algorithm, see Section \ref{sec:multi-controll}. Crucially, both algorithms fall within the complexity class of $\mathcal{O}(\poly(n))$. Based on this analysis, we conclude that the overall algorithm maintains a complexity of $\mathcal{O}(\poly(n))$.

%
 \subsection{Quantum Fourier Transform \label{sec:qft}}
%
We have introduced the classical Discrete Fourier Transform (DFT) of a vector~${\vec q} \in \mathbb C^N$ in Section~\ref{sec:fourierDiscretization}. The DFT can be expressed compactly in matrix form as 
\begin{equation} \label{eq:dft}
	\hat{\vec q} = {\vec F}_N {\vec q} \, ,
\end{equation}
where~${\vec F}_N \in \mathbb C^{N \times N}$ is the Fourier matrix
\begin{equation}
	{\vec F}_N = \frac{1}{\sqrt N} 
	\begin{pmatrix}
		 & \vdots & \\
	\dotsc & \omega_N^{jk} & \dotsc \\ 
	 	& \vdots	&  
	\end{pmatrix} \, , 
\end{equation}
and~$\omega_N$ is the N-th root of unity 
\begin{equation}
	\omega_N = e^\frac{2 \pi i }{N} \, .
\end{equation}
It is straightforward to show that the Fourier matrix~${\vec F}_N$ is unitary, i.e.~${\vec F}_N^{-1} = {\vec F}_N^\dagger $. Consequently, the DFT can be readily implemented on a quantum computer. 

In a quantum system with~$n$ qubits, the two vectors~$\hat{\vec q} \in \mathbb C^N $ and~${\vec q} \in \mathbb C^N$ are encoded into two state vectors $\ket q$ and $\widehat {\ket q}$, respectively, using $n$-qubits with~$N=2^n$. To emphasise the switch from DFT to QFT, we rewrite~\eqref{eq:dft} in bra-ket notation as 
\begin{equation} \label{eq:qft}
	\widehat {\ket q} = F_N \ket q \quad \Rightarrow \quad \sum_{j=0}^{N-1} \widehat q_j \ket j = F_N \sum_{k=0}^{N-1} q_k \ket k \, . 
\end{equation}
As always, in order to implement QFT on a quantum computer the unitary matrix~$F_N$ must be decomposed into elementary gates acting on one or two qubits at a time. In coming up with such a decomposition, it is, as usual, convenient to consider only the mapping of individual basis vectors, i.~e.,  
\begin{equation} \label{eq:qft_basis_transf}
	F_N {\ket k} = \frac{1}{\sqrt {N}} \sum_{j=0}^{N-1} \omega_N^{kj} \ket j = \frac{1}{\sqrt {N}} \sum_{j=0}^{N-1} e^{\frac{2 \pi i}{N} k j} \ket j \, .
\end{equation}
Subsequently, the resulting decomposition and quantum circuit can be applied to the entire state vector~$\ket q$ with no further changes,  owing to the quantum superposition principle. 

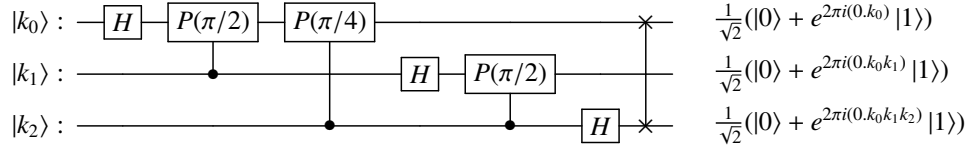
\begin{figure}[]
\centering 
\scalebox{1.0}{
\Qcircuit @C=1.0em @R=0.5em @!R { \\
	 	\lstick{\ket{{k}_{0}} : } & \gate{H} & \gate{P(\pi/2)} & \gate{P(\pi/4)} & \qw & \qw & \qw & \qswap \qwx[2] & \qw &  \rstick{ \frac{1}{\sqrt 2} ( \ket 0 + e^{2 \pi i ( 0.k_0 ) } \ket{1} ) }\\
		 \lstick{\ket{{k}_{1}} : } & \qw & \ctrl{-1} & \qw & \gate{H} & \gate{P(\pi/2)} & \qw & \qw & \qw & \rstick{ \frac{1}{\sqrt 2} ( \ket 0 + e^{2 \pi i ( 0.k_0 k_1 ) } \ket{1} ) } \\
  	 	\lstick{\ket{{k}_{2}} : } & \qw & \qw & \ctrl{-2} & \qw & \ctrl{-1} & \gate{H} & \qswap & \qw &  \rstick{ \frac{1}{\sqrt 2} ( \ket 0 + e^{2 \pi i ( 0.k_0 k_1 k_2 ) } \ket{1} )}     \\
\\ }}
\caption{Quantum Fourier Transform. One Hadamard gate~$H$ and a variable number of phase gates $P(\cdot)$ are applied to each qubit. For instance, providing the circuit with the input~$\ket k = \ket 1 \ket 1 \ket 0$ so that~$0.k_2 = 0$, $0.k_1 k_2 = 1/2$ and $0.k_0 k_1 k_2 = 3/4$, cf.~\eqref{eq:binary_frac} , it will yield the output state~\mbox{$ 1/(2 \sqrt{2})  (\ket 0 +\ket 1) (\ket 0 - \ket 1)  (\ket 0 - i \ket 1) $}. After evaluating the Kronecker products the result expressed in matrix form reads~\mbox{$(1, -i, -1, +i, 1, -i, -1, i )$}. \label{fig:qft_circuit} }
\end{figure}

We choose, by way of example,~$n=3$ to illustrate the decomposition of~$F_N$ into elementary one and two-qubit gates. Again, we make use of the binary representation of indices 
\begin{equation}
	k = k_0 k_1 k_2 = k_0 2^2+ k_1 2^1 + k_2 2^0, \quad j = j_0 j_1 j_2 = j_0 2^2 + j_1 2^1 +j_2 2^0 \, , 
\end{equation}
cf.~\eqref{eq:multiQlabeling}, and introduce the binary fraction notation 
\begin{equation} \label{eq:binary_frac}	
	\frac{k}{2^{n=3}} = 0.k_0 k_1 k_2 =  \frac{ k_0}{2^{1}} + \frac{ k_1}{2^{2}} + \frac{k_2}{ 2^{3}} \, .
\end{equation}
Furthermore, noting that~$\ket k = \ket{k_0 k_1 k_2}$ and~$\ket j = \ket{j_0 j_1 j_2}$, eq.~\eqref{eq:qft_basis_transf} can be rewritten as
\begin{align}
\begin{split} \label{eq:qft_product_form}
	F_N {\ket {k_0 k_1 k_2}} &= \frac{1}{\sqrt{2^3}} \sum_{j_0=0}^1 \sum_{j_1=0}^1 \sum_{j_2=0}^1 \left ( e^{2 \pi i ( 0.k_0 k_1 k_2 ) j_{0} 2^2 } \ket{j_{0}} \right ) \left ( e^{2 \pi i ( 0.k_0 k_1 k_2 ) j_1 2^1 } \ket{j_1} \right )  \left ( e^{2 \pi i ( 0.k_0 k_1 k_2 ) j_2 2^0} \ket{j_2} \right ) \, , \\ 
	&= \frac{1}{\sqrt{2^3}} \left (\ket 0 + e^{2 \pi i ( 0.k_0 k_1 k_2 ) 2^2  } \ket{1} \right ) \left ( \ket 0 + e^{2 \pi i ( 0.k_0 k_1 k_2 ) 2^1 } \ket{1} \right )  \left ( \ket 0 + e^{2 \pi i ( 0.k_0 k_1 k_2 ) 2^0 } \ket{1} \right ) \, .
\end{split}	
\end{align}
The exponentials can be further simplified since~\mbox{$e^{2 i \pi m} = \cos (2 \pi m) + i \sin(2  \pi m) \equiv 1 \; \forall m \in \mathbb Z^+$}, i.~e.,
\begin{equation}
	e^{2 \pi i ( 0.k_0 k_1 k_2 ) j_{0} 2^{2}} = e^{2 \pi i (0. k_2 ) j_{0} } , \quad e^{2 \pi i ( 0.k_0 k_1 k_2 ) j_{1} 2^{1}} = e^{2 \pi i (0. k_1 k_2 ) j_{1} } , \quad e^{2 \pi i ( 0.k_0 k_1 k_2 ) j_{2} 2^{0}} = e^{2 \pi i (0.k_0 k_1 k_2 ) j_{2} } \, ,
\end{equation}
Finally, a basis vector~$\ket k$ is mapped using QFT as
\begin{align}
\begin{split} \label{eq:qft_product_form}
	F_N {\ket {k_0 k_1 k_2}} = \frac{1}{\sqrt{2^3}} \left (\ket 0 + e^{2 \pi i ( 0.k_2 )  } \ket{1} \right ) \left ( \ket 0 + e^{2 \pi i ( 0.k_1 k_2 ) } \ket{1} \right )  \left ( \ket 0 + e^{2 \pi i ( 0.k_0 k_1 k_2 ) } \ket{1} \right ) \, .
\end{split}	
\end{align}
Each of the three factors correspond to a single qubit. We obtain each of the factors by successively mapping a basis~$\ket k = \ket {k_0 k_1 k_2}$ using the Hadamard and phase gates~$H$ and~$P(\theta)$, respectively, introduced in \eqref{eq:hadamard_phase}. For instance, focusing on the qubit~$\ket{k_0}$, the application of the Hadamard gate yields 
\begin{equation}
	H \colon \ket{k_0} \mapsto \frac{1}{\sqrt 2} \left ( \ket 0 + (-1)^{k_0} \ket 1\right ) = \frac{1}{\sqrt 2} \left ( \ket 0 + e^{\frac{2 \pi i}{2} {k_0} } \ket 1\right ) = \frac{1}{\sqrt 2} \left ( \ket 0 + e^{2 \pi i {(0.k_0)} } \ket 1\right ) \, .
\end{equation}
Subsequent application of the controlled phase gate~$P(\pi/2)$ with~$\ket{ k_1}$ as the control qubit gives
\begin{equation}
	\frac{1}{\sqrt 2} \left ( \ket 0 + e^{\frac{\pi}{2} i k_1} e^{2 \pi i {(0.k_0)} } \ket 1\right ) = \frac{1}{\sqrt 2} \left ( \ket 0 + e^{2 \pi i {(0.k_0 k_1)} } \ket 1\right ) \, . 
\end{equation}
Note that~$k_1 \in \{0, \, 1 \}$ and~$P(\pi/2)$ is only applied when~$k_1=1$. One last application of the controlled phase gate~$P(\pi/4)$ yields the last factor in~\eqref{eq:qft_product_form}, i.~e., 
\begin{equation}
	\frac{1}{\sqrt 2} \left ( \ket 0 + e^{\frac{\pi}{4} i k_2} e^{2 \pi i {(0.k_0 k_1)} } \ket 1\right ) = \frac{1}{\sqrt 2} \left ( \ket 0 + e^{2 \pi i {(0.k_0 k_1 k_2)} } \ket 1\right ) \, .
\end{equation} 
The second and first factors can be implemented in a similar way. The final circuit diagram is depicted in Figure~\ref{fig:qft_circuit}. Although the output of the circuit is correct, the values are reversed in comparison to the standard DFT and it is necessary to apply a swap gate to the qubits~$\ket{k_2}$ and~$\ket {k_0}$.

Finally, it is straightforward to set up a similar circuit for the inverse Fourier matrix~$F_N^{-1}$. To this end, the quantum circuit must be run in reverse and each gate replaced with its inverse, or its conjugate transpose since all the gates are unitary. 

\paragraph{Complexity} It is well known that the QFT circuit requires $n$ Hadamard and $n(n+1)/2$ controlled Phase gates \cite{ikeAndMike} . Therefore, the overall complexity of the QFT circuit is $\mathcal{O}(n^2)$. 

%
\subsection{Amplitude swap \label{sec: base swap}}
\begin{figure} 
\centering
\scalebox{1.0}{
\Qcircuit @C=1.2em @R=0.5em @!R { \\
	 	  \lstick{ \ket {k_0} :  } & \ctrlo{1}& \ctrl{1}  \barrier[0em]{2}  & \qw & \targ \barrier[0em]{2} & \qw & \ctrl{1} &  \ctrlo{1}  & \qw & \qw\\
	 	   \lstick{ \ket{k_1} :  } & \ctrl{1} & \ctrl{1}  & \qw  & \qw& \qw  & \ctrl{1}   & \ctrl{1} & \qw & \qw\\
	 	 \lstick{ \ket 0 :  }  & \targ & \targ & \qw & \ctrl{-2}  & \qw & \targ   & \targ & \qw & \qw\\
\\ }}
\caption{Quantum circuit for swapping the components~$q_1$ and~$q_3$ of a state vector~\mbox{$\ket q = q_0 \ket{00}+ q_1 \ket{01} + q_2 \ket{10}+ q_3 \ket{11}$}. The two dashed helper lines correspond to the the three distinct steps discussed in the main text. }
\label{fig:base_swap circ}
\end{figure}
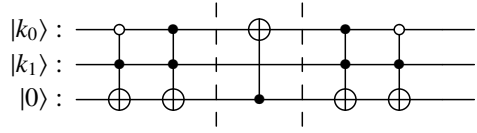

%
The swapping of two amplitudes of a state vector is a useful operation in quantum computing. We consider the state vector
\begin{equation}
	\ket q = \sum_{k=0}^{2^n-1} q_k \ket k = \sum_{k=0, k\neq k', k \neq k''}^{2^n-1} q_k \ket k  + q_{k'}  \ket{k'} + q_{k''}  \ket{k''}      \, ,
\end{equation}
where~$q_{k'} $ and~$ q_{k''} $ are the amplitudes to be swapped. The first step in implementing the swap is to tag the two amplitudes in question, which can be accomplished by introducing an ancillary qubit, cf. Figure~\ref{fig:base_swap circ}. The state of the ancilla is set using multi-controlled~$CNOT$ gates with their controls determined according to the binary representations of~$k'$ and~$k''$, so that, 
\begin{equation}
	 \sum_{k=0, k\neq k', k \neq k''}^{2^n-1} q_k \ket k \ket 0  + q_{k'}  \ket{k'} \ket 0 + q_{k''}  \ket{k''}  \ket 0   \mapsto  \sum_{k=0, k\neq k', k \neq k''}^{2^n-1} q_k \ket k \ket 0  + q_{k'}  \ket{k'} \ket 1 + q_{k''}  \ket{k''}  \ket 1  \, .
\end{equation}
Subsequently, in the second step the tagged amplitudes are swapped using~$CNOT$ gates conditioned on the ancillary yielding the state
\begin{equation}
	  \sum_{k=0, k\neq k', k \neq k''}^{2^n-1} q_k \ket k \ket 0  + q_{k'}  \ket{k'} \ket 1 + q_{k''}  \ket{k''}  \ket 1 \mapsto  \sum_{k=0, k\neq k', k \neq k''}^{2^n-1} q_k \ket k \ket 0  + q_{k''}  \ket{k'} \ket 1 +  q_{k'}  \ket{k''}  \ket 1  \, .
\end{equation}
In the final step, the state of the ancilla qubit is reset to its initial state~$\ket 0$, which can be accomplished by applying (in reverse order) the same set of multi-controlled $CNOT$ gates used in the first step. This final transformation of the state vector reads
\begin{equation}
	  \sum_{k=0, k\neq k', k \neq k''}^{2^n-1} q_k \ket k \ket 0  + q_{k''}  \ket{k'} \ket 1 +  q_{k'}  \ket{k''}  \ket 1   \mapsto \sum_{k=0, k\neq k', k \neq k''}^{2^n-1} q_k \ket k \ket 0  + q_{k''}  \ket{k'} \ket 0 +  q_{k'}  \ket{k''}  \ket 0   .
\end{equation}
An example circuit for swapping the components~$q_1$ and~$q_3$, i.e.~$k'=1 \equiv 01$ and~$k''=3 \equiv 11$, of a state vector~\mbox{$\ket q \in \mathbb C^{4}$} is shown in Figure~\ref{fig:base_swap circ}. 

\paragraph{Complexity} The multi-controlled~$CNOT$ gate is key to the implementation of the introduced amplitude-swap circuit.  The circuit requires~$4$ multi-controlled~$CNOT$ gates so that we need~$8n-12$ Toffoli gates in total, see Section~\ref{sec:multi-controll}.  In addition, the circuit requires in the second step of the algorithm up to~$n$ $CNOT$ gates. We can conclude that for a state vector with~$N=2^n$ components the number of universal gates~$\{CNOT, \, U_3 \}$ will scale as~$\mathcal{O} ( n)$, or $\mathcal{O} ( \log \, (N))$.

%
\section{Quantum computational homogenisation}
%
We turn to the implementation of the periodic homogenisation approach introduced in Section~\ref{sec:compHomog} on a quantum computer using the quantum algorithms described in Section~\ref{sec:algorithms}. As detailed in Section~\ref{sec:fourierDiscretization}, we approximate all periodic fields using band-limited Fourier interpolation. Consequently, we solve linear differential equations by solving a set of decoupled algebraic equations in Fourier space. The transformations between the physical and Fourier spaces are carried out using the QFT. For ease of presentation, we first consider the solution of periodic Poisson problems in 1D and 2D and, subsequently, the homogenisation problem on an RVE. 

All the proposed quantum algorithms are implemented using Python-based Qiskit SDK~\cite{Qiskit} and executed on a noiseless simulator. In all cases, we verify the correctness of the implementation by comparing the quantum solution encoded in the output state vector with analytical and classical numerical solutions.
%
\subsection{Periodic Poisson problems}

We recall that Algorithm~\ref{alg:DFT} sets forth a sequence of Poisson problems, one for every iteration. We therefore begin by discussing the implementation of that class of problems.
%
\subsubsection{One-dimensional case}
%
We seek the solution of the one-dimensional periodic Poisson problem
\begin{equation} \label{eq:oneDex}
	- \frac{\D ^2 v (x)}{\D x^2}  = f(x) \, , 
\end{equation}
where the solution~$v(x)$ and the source~$f(x)$ are $L$-periodic, i.~e.~\mbox{$v(x) = v(x+L)$} and \mbox{$f(x)=f(x+L)$}. In order to solve this problem, it suffices to consider the domain~$\Omega=(0, \, L) \in \mathbb R$ and discretise it using a uniform grid of~$N$ cells of size~$h$. For the sake of simplicity, we have chosen $L=1$ in this example. A grid point with the index~$k \in \{0, \, 1, \, \dotsc, \, N-1\}$ has coordinate~$x_k = k h $ and source value~$ f_k = f(x_k)$. The source values at the~$N$ grid points are collected in the ket
\begin{equation}
	\ket f = \sum_{k=0}^{N-1} f_k \ket k \, ,
\end{equation}
with QFT 
\begin{equation}
	 \sum_{j=0}^{N-1} \hat f_j \ket j = F_N \sum_{k=0}^{N-1} f_k \ket k \, ;
\end{equation}
see Section~\ref{sec:qft}. Following the derivation in Sections~\ref{sec:periodicSol} and~\ref{sec:fourierDiscretization}, the solution of the periodic Poisson problem in Fourier space follows as 
\begin{equation}
	\hat v_{j} = \left ( \frac{L}{2 \pi r(j)} \right )^2 \hat f_{j} \, , 
\end{equation}
where the relabelling function~$r(j)$ is defined in~\eqref{eq:componentsToFrequencies}. An inverse QFT finally gives the solution in real space as
\begin{equation}
	\ket v = F_N^\dagger \widehat {\ket v } = F_N^\dagger \sum_{j=0}^{N-1} \hat v_j \ket j \, .
\end{equation}
Furthermore, the band-limited interpolation~\eqref{eq:fourierApprox} gives the approximate solution as
\begin{equation}
	v^h (x) = \frac{1}{N} \sum_{l=0}^{N-1} \hat v_{l} e^{i \frac{2 \pi r(l)}{L}  x} \, .
\end{equation}
at every $x \in [0,L]$. 

\begin{figure}[]
\centering
\scalebox{1.0}{
\Qcircuit @C=1.0em @R=1.0em @!R {
	\lstick{\ket{k_0} = \ket 0 :} & \multigate{2}{U_I ( \ket f ) } & \multigate{2}{ F_N} & \multigate{3}{ U_P \left (  \frac{L^2}{ (2 \pi r(k))^2} \right )} & \multigate{2}{F_N^{\dagger}} & \qw & \qw\\
	\lstick{\ket{k_1} = \ket 0 :} & \ghost{U_I( \ket f ) } & \ghost{F_N} & \ghost{ U_P \left ( \frac{L^2}{ \left ( 2 \pi r(k) \right )^2}  \right ) } & \ghost{F_N^{\dagger}} & \qw & \qw\\
	\lstick{\ket{k_2} = \ket 0:} & \ghost{U_I ( \ket f ) } & \ghost{F_N} & \ghost{U_P \left (  \frac{L^2}{ (2 \pi r(k))^2} \right )} & \ghost{F_N^{\dagger}} & \qw & \qw\\
	\lstick{ \ket 0:} & \qw & \qw & \ghost{U_P \left (  \frac{L^2}{ (2 \pi r(k))^2} \right )} & \qw & \qw & \qw\\
}}
\caption{Quantum circuit for solving the one-dimensional periodic Poisson problem discretised with a uniform grid of~$N=8$ cells. Spatial and Fourier coefficients at the $N=8$ grid points are encoded as the amplitudes of a quantum state of the~$n=3$ qubits~$\ket{k_0} \ket{k_1} \ket{k_2}$. The left most gate~$U_I$ encodes the source~$\ket f$ as a quantum state. The second gate applies QFT to determine~$\widehat{\ket f}$. In the Fourier space the Poisson problem is solved by scaling the Fourier components of~$\widehat{\ket f}$ with the the third gate~$U_P$. The last gate applies the inverse QFT to obtain the solution~$\ket v$. The unnamed last qubit is an ancillary needed internally by~$U_P$ as discussed in Section~\ref{sec:polyEncode}. }
\label{fig:1DPoissonQcirc}
\end{figure}
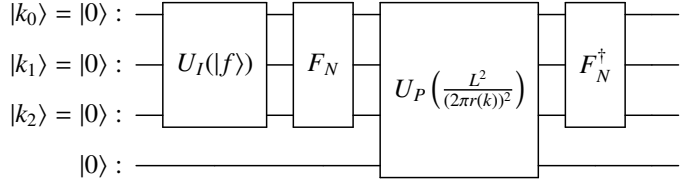

The proposed quantum circuit for solving the one-dimensional Poisson problem is depicted in Figure~\ref{fig:1DPoissonQcirc} for~$N=8$ grid cells. The circuit has been implemented using the Qiskit SDK. It has one ancillary qubit and three qubits~$\ket k \equiv \ket {k_0 k_1 k_2} $ for representing the source~$\ket f$ and solution~$\ket v$. All qubits are initialised in the state~$\ket 0$. The first unitary~$U_I$ prepares a quantum state representing the components of the source~$\ket f$. $U_I$ can be chosen using the state preparation technique introduced in Section~\ref{sec:statePrepEF}, or one of the many other known approaches. The unitaries~$F_N$ and~$F_N^{\dagger}$ apply the QFT and its inverse. Finally the unitary~$U_P$ multiplies the Fourier coefficients by $L^2/(2 \pi r(l))^2$ . 

The sequence of mappings implemented by the circuit with~$n=3$ qubits can be summarised as follows:
\begin{subequations} \label{eq:1DPoissonMappings}
\begin{align}
	U_I \left ( \ket f \right ) \otimes I  & \colon \quad  \ket 0^{\otimes 3} \ket 0  \mapsto \sum_{k=0}^7 f_k \ket k \ket 0 \, ,  \\ 
	F_N \otimes I & \colon \quad  \sum_{k=0}^7 f_k \ket k \ket 0 \mapsto  \sum_{j=0}^7 \hat f_j \ket j \ket 0 \, , \\ 
	U_P \left ( \frac{L^2}{ \left ( 2 \pi r(j) \right )^2}  \right ) & \colon  \quad \sum_{j=0}^7 \hat f_j \ket j \ket 0  \mapsto \textit{Junk} + \sum_{j=0}^7  \left ( \frac{L}{2 \pi r(j)} \right )^2 \hat f_j \ket j \ket 1 \, , \\ 
	F_N^{\dagger} \otimes I & \colon \quad  \textit{Junk} + \sum_{j=0}^7  \hat v_j \ket j \ket 1  \mapsto \textit{Junk} + \sum_{k=0}^7 v_k \ket k \ket 1 \, . 
\end{align}
\end{subequations}
In this implementation, all the irrelevant terms as regards the final state are denoted as~$Junk$. 

\begin{figure} 
\centering
\includegraphics[width=0.8\textwidth]{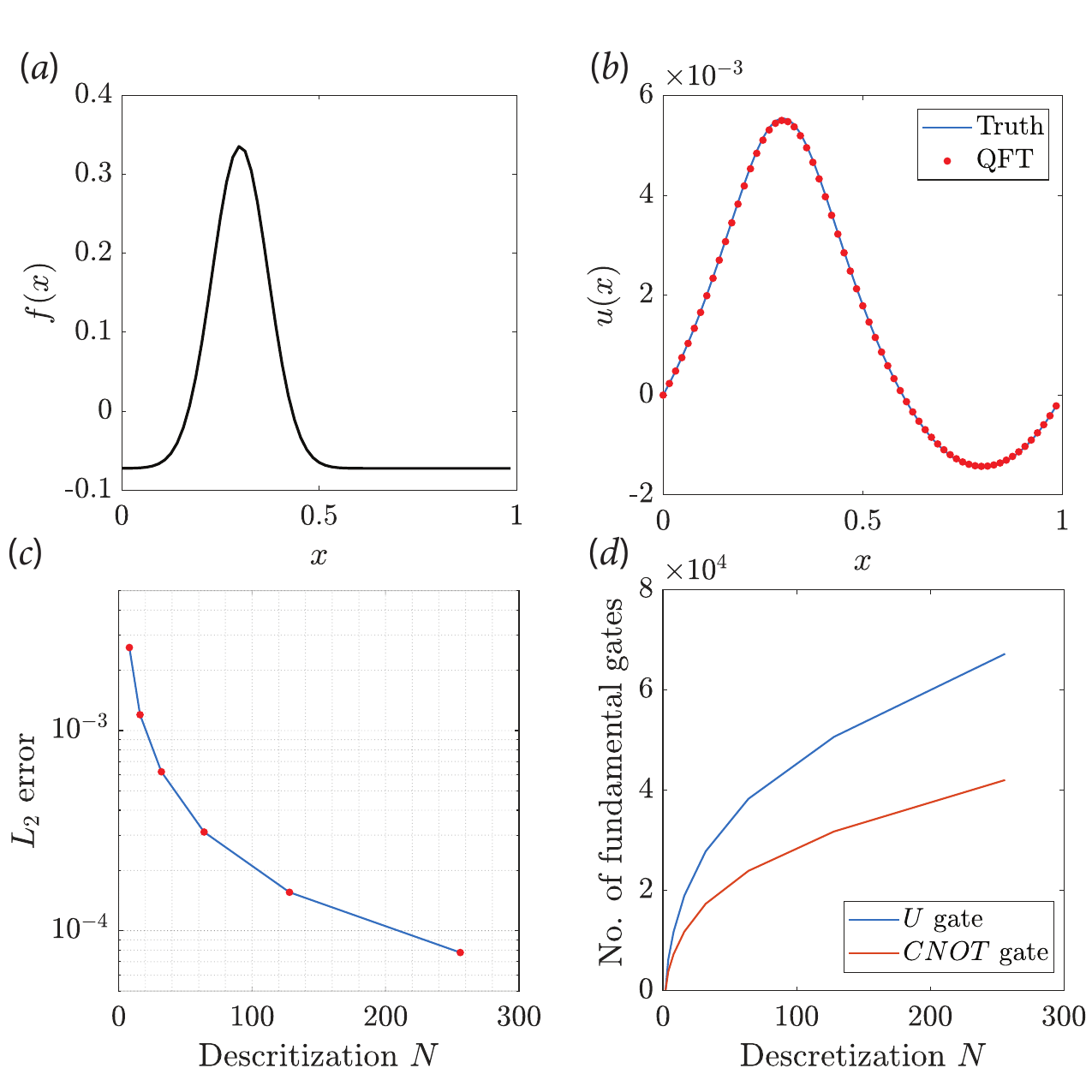}
\caption{One-dimensional periodic Poisson problem. (a) Source function $f(x)$. (b) Comparison of the quantum solution on a grid with $64$ cells and the exact (analytical solution). (c) Convergence of the $L_2$ error. (d) Scaling of the total number of~$U_3$ and~$CNOT$ gates. 
\label{fig:1DPoisson}}
\end{figure}

We assess the convergence properties and the computational complexity of the algorithm by applying it to the source 
\begin{equation}
	f(x) = \exp \left (- \frac{( x-\alpha_0)^2}{\alpha_1^2} \right ) + \alpha_2 \, ,
\end{equation}
where the three parameters are chosen as~$\alpha_0 = 0.3$, $\alpha_1 = 0.1$, $\alpha_2 = -0.1772$. This function is depicted in Figure~\ref{fig:1DPoisson}(a) and is approximated as periodic. These parameters are chosen such that $\int_0^L f(x)dx$ = 0 to ensure the solution of the problem is well defined. The respective analytical solution is 
\begin{equation} 
	v (x) = \frac{\alpha_1 \sqrt \pi}{2}\Bigl ((x-\alpha_0)\erf \left (\frac{x-\alpha_0}{\alpha_1} \right ) + \frac{\alpha_1}{\sqrt{\pi}} \exp \left (- \frac{(x-\alpha_0)^2}{\alpha_1^2} \right ) \Bigr )+\frac{\alpha_2x^2}{2}+\alpha_3x+\alpha_4 \, . 
\end{equation}
The periodic boundary conditions~$u(0)=u(1) = 0$ determine the two integration constants as $\alpha_3 = 0.053$ and $\alpha_4 = -0.0266$. 

In Figure~\ref{fig:1DPoisson}(b) the analytic solution~$v(x)$ and the approximate solution~$v^h(x)$ obtained with quantum computing are compared, evincing an excellent agreement between the two. The convergence of the $L_2$-norm error for an increasing number of grid cells is shown in Figure~\ref{fig:1DPoisson}(c). The error is defined as 
\begin{equation}
	\eta = {\| \ket v - \ket { v^{ex} } \| }\, ,
\end{equation}
where~$\ket { v^{ex} } = \sum_{k=0}^{N-1} v (x_k) \ket k$ is the exact solution at the grid points. The error has a convergence order of~$1/2$, which is mainly dictated by the piecewise polynomial approximation in the unitary~$U_P( L^2 / (2 \pi r(l))^2 )$.

\bigskip

\paragraph{Complexity} Finally, we assess the computational complexity of the proposed algorithm by expressing the quantum circuit using only the single-qubit rotation gate $U_3$ and the two-qubit controlled gate $CNOT$. These two gates are universal in the sense that any arbitrary multi-qubit unitary can be composed using them. In quantum computing the process of expressing a given circuit in terms of a different gate set is referred to as \emph{transpiling} and can be automatically performed in Qiskit. The total number of~$U_3$ and~$CNOT$ gates as a function of the number of grid cells~$N$ is plotted in Figure~\ref{fig:1DPoisson}(c), clearly exhibiting the desired logarithmic scaling for both the~$U_3$ and~$CNOT$ gates in agreement with the expected computational complexity $\mathcal{O} \ (\poly (\log (N)))$. 

%
\subsubsection{Two-dimensional case}
%
Next, we consider the two-dimensional $L$-periodic Poisson problem 
\begin{equation}
	- \frac{\partial^2 v(\vec x)}{\partial x_0^2} - \frac{ \partial^2 v(\vec x)}{\partial x_1^2} = f(\vec x) \, .
\end{equation}
The solution and source fields~$v(\vec x)$ and~$f(\vec x)$ have periodicity~\mbox{$v(\vec x) = v(\vec x+ \vec e_0 L) = v(\vec x+ \vec e_1 L) $} and~$f(\vec x) = f(\vec x+ \vec e_0 L) = f(\vec x+ \vec e_1 L) $, where~\mbox{$\vec e_0 = ( 1 ,\, 0)^\trans$} and~\mbox{$\vec e_1 = ( 0 ,\, 1)^\trans$} are the standard basis vectors. 

We discretise the domain~$\Omega = (0, \, L)^2 \subset \mathbb R^2$ by uniformly partitioning it into ~$N \times N$ cells of size~$h$, where we choose $L=1$ for the sake of simplicity. Consequently, a grid point with multi-index~$\vec k = (k^0, \, k^1 )$, where $k^0, k^1 \in \{0, \, 1, \, \dotsc, \, N-1\}$, has the coordinates
\begin{equation}
	\vec x_{\vec k} = \begin{pmatrix} x_{0, \vec k} & x_{1, \vec k} \end{pmatrix}^\trans = \begin{pmatrix} x_{0, k^0 k^1} & x_{1, k^0 k^1} \end{pmatrix}^\trans \, , 
\end{equation}
and the source value~$f_{\vec k} = f(\vec x_{\vec k}) $. We collect the source values in the ket
\begin{equation}
	\ket f = \sum_{\vec k=0}^{N-1} f_{\vec k} \ket k = \sum_{k^0=0, k^1=0}^{N-1} f_{k^0 k^1} \ket {k^0 k^1 } \, .
\end{equation}
As in standard DFT, the two-dimensional QFT is the Kronecker product of two one-dimensional QFT's~$F_{0,N}$ and~$F_{1,N}$. Hence, we can compactly write 
\begin{equation}
 \widehat {\ket f} = (F_{0,N} \otimes F_{1,N}) \ket f \, ; 
\end{equation}
or in more in detail, 
\begin{equation}
	 \sum_{j^0=0, j^1=0}^{N-1} \hat f_{j^0 j^1} \ket {j^0 j^1 } = ( F_{0,N} \otimes F_{1,N}) \sum_{k^0=0, k^1=0}^{N-1} f_{k^0 k^1} \ket {k^0 k^1 } \, . 
\end{equation}
Here,~$F_{0,N}$ and~$F_{1,N}$ denote the QFTs with respect to the first and second indices of the multi-index, respectively. Building on the periodic one-dimensional Poisson problem introduced before, the solution of the two-dimensional problem in Fourier space follows as
\begin{equation} \label{eq:scalingFourier2D}
\hat v_{j^0 j^1} = \frac{L^2}{4 \pi^2 \left (  r^2 \left (j^0 \right )+r^2\left (j^1 \right ) \right )}  \hat f_{j^0 j^1} \, ;
\end{equation}
where~$r(j^0)$ and~$r(j^1)$ are defined in~\eqref{eq:componentsToFrequencies}. Finally, the inverse QFT of the computed Fourier coefficients yields the solution at the grid points, i.~e., 
\begin{equation}
	\ket v = \left ( F_{0,N}^\dagger \otimes F_{0,N}^\dagger 
 \right ) \widehat {\ket v } =  \left ( F_{0,N}^\dagger \otimes F_{0,N}^\dagger \right )  \sum_{j^0=0 j^1=0}^{N-1} \hat v_{j^0 j^1} \ket {j^0 j^1}  =
  \sum_{k^0=0 k^1=0}^{N-1}  v_{k^0 k^1} \ket {k^0 k^1} 
 \, .
\end{equation}

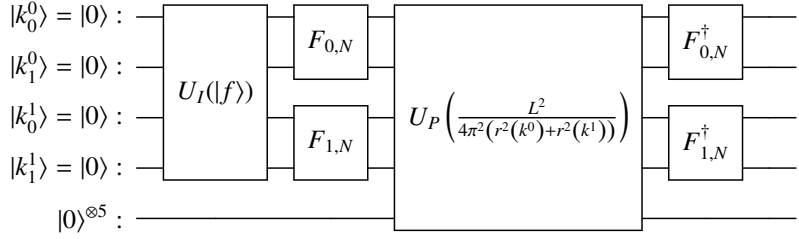
\begin{figure} 
\centering
\scalebox{1.0}{
\Qcircuit @C=1.0em @R=1.0em @!R { \\
	 	 \lstick{ \ket{k_0^0} = \ket 0 : } & \multigate{3}{U_I(\ket f) } & \multigate{1}{F_{0,N}} & \multigate{4}{U_P\left (   \frac{L^2}{4 \pi^2 \left (  r^2 \left (k^0 \right )+r^2\left (k^1 \right ) \right )}  \right )} & \multigate{1}{F_{0,N}^\dagger} & \qw & \qw\\
	 	 \lstick{ \ket{k_1^0} = \ket 0 : } & \ghost{U_I(\ket f) } & \ghost{F_{0,N}} &    \ghost{U_P\left (  \frac{L^2}{4 \pi^2 \left (  r^2 \left (k^0 \right )+r^2\left (k^1 \right ) \right )}  \right )}  & \ghost{F_{0,N}^\dagger} & \qw & \qw\\
	 	 \lstick{ \ket{k_0^1} = \ket 0 :  } & \ghost{U_I(\ket f) } & \multigate{1}{F_{1,N}} &  \ghost{U_P\left (  \frac{L^2}{4 \pi^2 \left (  r^2 \left (k^0 \right )+r^2\left (k^1 \right ) \right )}   \right )}   & \multigate{1}{F_{1,N}^\dagger} & \qw & \qw\\
	 	 \lstick{ \ket{k_1^1} = \ket 0 : } & \ghost{U_I(\ket f) } & \ghost{F_{1,N}} &  \ghost{U_P\left (  \frac{L^2}{4 \pi^2 \left (  r^2 \left (k^0 \right )+r^2\left (k^1 \right ) \right )}  \right )}  & \ghost{F_{1,N}^\dagger} & \qw & \qw\\
	 	 \lstick{ \ket {0}^{\otimes 5} :} & \qw & \qw &  \ghost{U_P\left (   \frac{L^2}{4 \pi^2 \left (  r^2 \left (k^0 \right )+r^2\left (k^1 \right ) \right )} \right )}  & \qw & \qw & \qw\\
\\ }}
\caption{ Quantum circuit for solving the two-dimensional periodic Poisson problem discretised with a uniform grid of $N \times N = 4 \times 4 $ cells. Spatial and Fourier coefficients at the~$N^2 = 16$ grid points are encoded as the amplitudes of a quantum state of the $n = 4$ qubits~\mbox{$ \ket { k^0_0}  \ket{k^0_1 } \ket {k^1_0} \ket{ k^1_1}$}. The left most gate~$U_I$ encodes the source $\ket f $ as a quantum state. The second layer of gates~$F_{0,N}$ and~$F_{1,N}$ apply simultaneously the QFT with respect to the multi-indices~$k^0$ and~$k^1$. In the Fourier space the Poisson problem is solved by scaling the Fourier components of~$\widehat{\ket f} $ with the gate~$U_P$. The last two gates apply the inverse QFT to obtain the solution~$\ket v$. The unnamed last five qubits are ancillaries needed by~$U_P$, see Sections~\ref{sec:polyEncode} and~\ref{sec:encode_function}. 
}
\label{fig:2DPoissonCirc}
\end{figure}

The quantum circuit for solving the two-dimensional Poisson problem can be designed by closely following the one-dimensional construction introduced in the previous section. As an example, in Figure~\ref{fig:2DPoissonCirc} the circuit for a grid with~$4 \times 4$ cells is shown. In mapping the source~$\ket f$ and other kets to qubits we make use of the following expansion of the multi-indices,
\begin{equation}
	\ket {\vec k} = \ket { k^0} \ket {k^1} = \ket { k^0_0 k^0_1 } \ket {k^1_0 k^1_1} \equiv  \ket { k^0_0}  \ket{k^0_1 } \ket {k^1_0} \ket{ k^1_1} \, .
\end{equation}
After the second equality sign we express the two integers~$k^0, k^1 \in \mathbb \{0, \, 1, \, 2, \, 3 \}$ as binaries. In Figure~\ref{fig:2DPoissonCirc} the unitary~$U_I$ for state preparation and the unitaries~$F_{0,N}$ and~$F_{1,N}$ and their inverses for the QFT follow along the lines of the unitaries in the circuit for the one-dimensional problem in Figure~\ref{fig:1DPoissonQcirc}. The main difference concerns the unitary~$U_P$ that encodes a bivariate polynomial for implementing~\eqref{eq:scalingFourier2D}. Importantly, owing to quantum parallelism the QFT's with respect to the two multi-indices can be applied simultaneously as evident from the quantum circuit. The efficiency implications of this parallelism for high-dimensional problems are evident and far reaching.

\begin{figure} 
\centering
\includegraphics[width=0.8\textwidth]{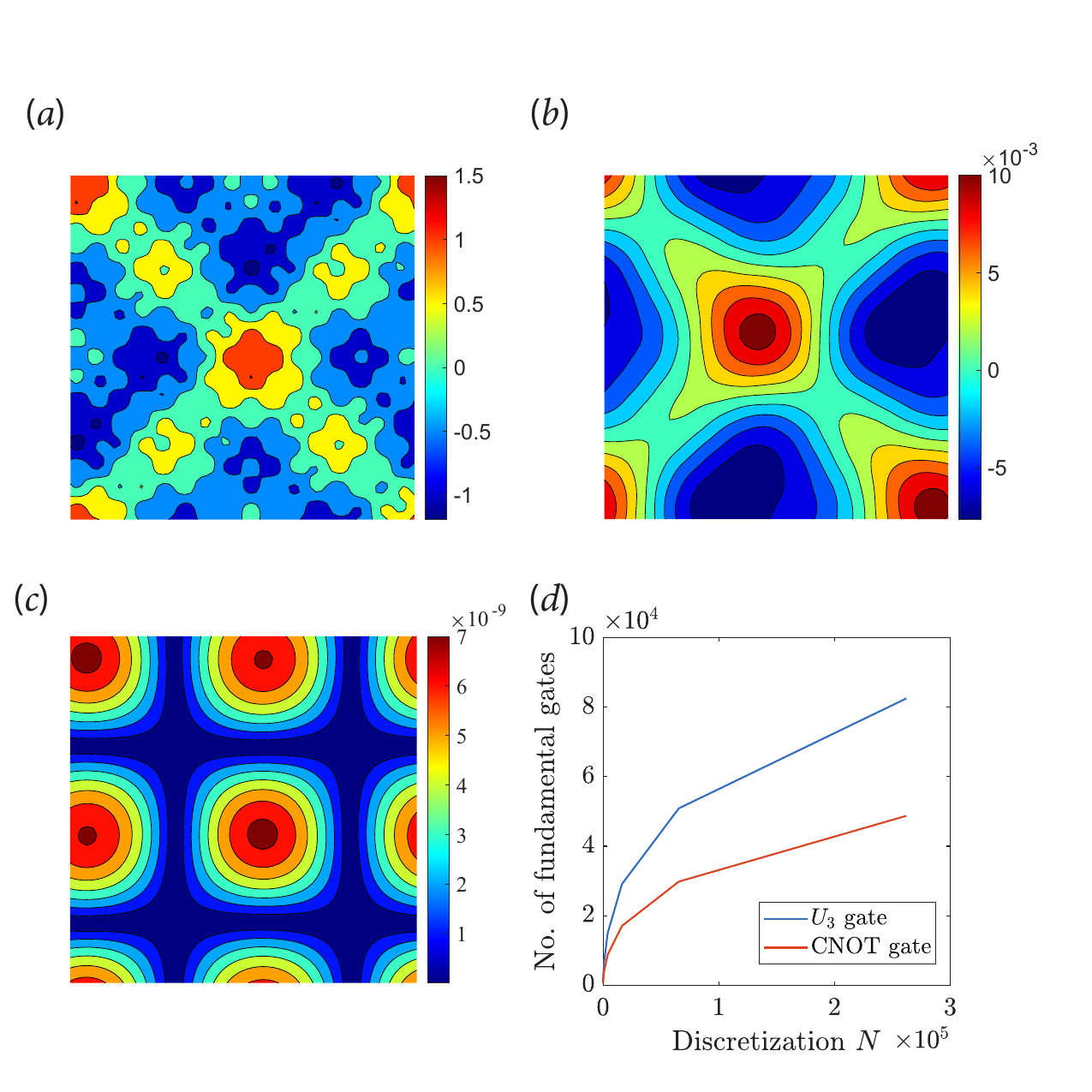}
\caption{Two-dimensional periodic Poisson problem. (a) Source function $f(x)$. (b) Quantum solution obtained with a grid with $64 \times 64$ cells. (c) Absolute error of the quantum solution in (b) and the true solution. (d) Scaling of the total number of~$U_3$ and~$CNOT$ gates.
\label{fig:2DPoisson} }
\end{figure}

As a concrete example, we consider a grid of size $N\times N = 64 \times 64$ and the source 
\begin{equation} 
	f(x_0, \, x_1) = \sum_{l=0}^{M} \alpha_{0,l} \sin(\alpha_{1,l} \pi x_0 + \alpha_{2,l}) \sin(\alpha_{3,l} \pi x_1 + \alpha_{4,l}) \, , 
\end{equation}
    where $\alpha_{0,l}, \, \alpha_{2,l}, \, \alpha_{4,l} \in \mathbb{R}$ and $\alpha_{1,l}, \, \alpha_{3,l} \in \mathbb{Z}$ are five random parameters. The real parameters are sampled from a uniform probability measure \mbox{$\mathcal D_{R} (-1, \, 1)$} and the integer parameters from a uniform probability measure $\mathcal D_{Z} (-20, \, 20)$. Similar to the 1D case, we have chosen the parameters such that $\int_{\Omega}f(x_0,x_1) d\Omega =0$. An example~$f(x_0, \, x_1)$ with~$M=3$ is shown in Fig \ref{fig:2DPoisson}(a). The corresponding quantum solution~$\ket v$ is depicted in Figure \ref{fig:2DPoisson}(b), and the distribution of  absolute error $||v\rangle-|v^{ex}\rangle|$ of quantum solution and true solution $|v^{ex}\rangle$ is shown in \ref{fig:2DPoisson}(c). The true solution is obtained by solving the same problem using a classical solver on a fine grid of size~$1024\times1024$. We observe the high accuracy of the proposed quantum algorithm with the largest absolute error of the order of $10^{-8}$. 

\bigskip

\paragraph{Complexity} Finally, we assess the complexity of the algorithm by counting total number of~$U_3$ and CNOT gates as a function of the number of grid cells~$N^2$, see Figure~\ref{fig:2DPoisson}(d). The desired $\mathcal{O} ( \poly\log(N^2))$ scaling is again evident from the plot. 

%
\subsection{Solution of the unit cell problem}
%
We now return to the solution of the periodic homogenisation problem on a quantum computer. For simplicity, the RVE is one-dimensional with a domain~$\Omega = (0, \, L) \subset \mathbb R$ and piecewise constant shear modulus~$\mu(x)$ as depicted in Figure~\ref{fig:QRVE}(a). This problem represents the homogenisation of an elastic composite rod with a periodic Young's modulus~$\mu(x)$. As discussed in Section~\ref{sec:compHomog}, homogenisation aims to determine the stress field~$\sigma(x)$ for a prescribed scalar applied strain~$\overline \gamma$. We achieve this by solving a sequence of Poisson problems as outlined in Algorithm~\ref{alg:DFT}. In the following, we first introduce the quantum circuit for computing one iteration step and subsequently explain how to execute several steps. 
%
\subsubsection{Iterative step}
%
We closely follow Algorithm~\ref{alg:DFT} in devising the quantum circuit for solving one homogenisation iteration. Assuming that the RVE is discretised into~$N=2^n$ cells, the quantum state vector~\mbox{$\ket q$} for implementing the circuit has the structure
\begin{equation} \label{eq:statOneIterativeStep}
	\ket{q} = \underbrace{\ket{0}^{\otimes n}}_\text{field}
	\underbrace{\ket { a_{0}} \ket{ a_{1}}}_\text{controls} 
	\underbrace{\ket {b}}_\text{AS}
	\underbrace{\ket 0^{\otimes n}}_\text{ancillas} \, , 
\end{equation} 
where the first~$n$ qubits encode field vectors, like the shear modulus~$\ket \mu \in \mathbb R^N$ or strain $\ket \gamma \in \mathbb R^N$, the qubits~$\ket {a_0}$ and~$\ket {a_1}$ control the application of unitaries and the qubit $\ket b$ encodes applied strain~$\overline \gamma \in \mathbb R$. The last~$n$ qubits serve as helpers throughout the circuit.

In the initialisation step of Algorithm~\ref{alg:DFT}, we encode the prescribed applied strain (AS)~$\overline \gamma$ 
as an amplitude of the state vector~$\ket{q}^{(s=0)}$. 
The applied strain~$\overline \gamma$ is encoded as the amplitude of~$\ket b$. This is easily accomplished by applying the rotation~$R_Y(\theta)$ gate with $\theta = 2 \arcsin \left (\sqrt{N}\bar{\gamma} \right )$ to qubit~$\ket b$, so that the quantum state becomes
\begin{equation} \label{eq:psi1}
	\ket{q}^{(0)}_1 =\sqrt{N}\bar{\gamma} \underbrace{\ket{0}^{\otimes n}}_\text{field}
	\underbrace{\ket {0} \ket{0}}_\text{controls} 
	\underbrace{\ket {1}}_\text{AS}
	\underbrace{\ket 0^{\otimes n}}_\text{ancillas}
	+ \sqrt{1- \left (\sqrt{N}\bar{\gamma} \right )^2}
	\underbrace{\ket{0}^{\otimes n}}_\text{field}
	\underbrace{\ket {0} \ket{0}}_\text{controls} 
	\underbrace{\ket {0}}_\text{AS}
	\underbrace{\ket 0^{\otimes n}}_\text{ancillas} \, .
\end{equation} 
The predictor strain~$\ket {\gamma}^{(0)}$ can be encoded by applying a standard unitary~$U_I(\ket {\gamma}^{(0)})$ for state preparation when the AS qubit is in state~$\ket {b = 0}$. The resulting new state reads 
\begin{equation} \label{eq:psi2}
	\ket{q}^{(0)}_2 =\sqrt{N}\bar{\gamma} \underbrace{\ket{0}^{\otimes n}}_\text{field}
	\underbrace{\ket {0} \ket{0}}_\text{controls} 
	\underbrace{\ket {1}}_\text{AS}
	\underbrace{\ket 0^{\otimes n}}_\text{ancillas}
	+ \sqrt{1-\left (\sqrt{N}\bar{\gamma} \right )^2}
	\underbrace{ \sum_{k=0}^{2^n-1} \gamma_k^{(s)} \ket k}_\text{field}
	\underbrace{\ket {0} \ket{0}}_\text{controls} 
	\underbrace{\ket {0}}_\text{AS}
	\underbrace{\ket 0^{\otimes n}}_\text{ancillas} \, .
\end{equation} 

\begin{figure}
\centering
\scalebox{1.0}{
\Qcircuit @C=1.0em @R=0.2em @!R { \\
	 	\lstick{\ket{k_0} = \ket 0 :}  & \qw & \multigate{6}{U_P \left ( \frac{\mu(x_{k}) - \mu^0}{\sqrt{1- \left (\sqrt{N}\bar \gamma \right )^2}} \right ) } & \multigate{2}{F_N} & \multigate{6}{U_P \left (\frac{1}{\mu^0} \right )} & \multigate{6}{U_S} & \multigate{2}{F_N^\dagger} & \qw & \qw\\
	 	\lstick{\ket{k_1} = \ket 0 :} & \qw & \ghost{U_P \left ( \frac{\mu(x_{k}) - \mu^0}{\sqrt{1- \left (\sqrt{N}\bar \gamma \right )^2}} \right )} & \ghost{F_N} & \ghost{U_P \left (\frac{1}{\mu^0} \right )} & \ghost{U_S} & \ghost{F_N^\dagger} & \qw & \qw\\
	 	\lstick{\ket{k_2} = \ket 0:}  & \qw & \ghost{U_P \left ( \frac{\mu(x_{k}) - \mu^0}{\sqrt{1- \left (\sqrt{N}\bar \gamma \right )^2}} \right )} & \ghost{F_N} & \ghost{U_P \left (\frac{1}{\mu^0} \right )} & \ghost{U_S} & \ghost{F_N^\dagger} & \qw & \qw\\
	 	 \lstick{\ket{{a}_{0}}= \ket 0 : } & \qw & \ghost{U_P \left ( \frac{\mu(x_{k}) - \mu^0}{\sqrt{1- \left (\sqrt{N}\bar \gamma \right )^2}} \right )} & \qw & \ghost{U_P \left (\frac{1}{\mu^0} \right )} & \ghost{U_S} & \qw & \qw & \qw\\
	 	 \lstick{\ket{{a}_{1}} = \ket 0 : } & \qw & \ghost{U_P \left ( \frac{\mu(x_{k}) - \mu^0}{\sqrt{1- \left (\sqrt{N}\bar \gamma \right )^2}} \right )} & \qw & \ghost{U_P \left (\frac{1}{\mu^0} \right )} & \ghost{U_S} & \qw & \qw & \qw\\
	 	 \lstick{\ket{{b}} = \ket 0 : } & \gate{U_I(\bar \gamma)} & \ghost{U_P \left ( \frac{\mu(x_{k}) - \mu^0}{\sqrt{1- \left (\sqrt{N}\bar \gamma \right )^2}} \right )} & \ctrl{-3} & \ghost{U_P \left (\frac{1}{\mu^0} \right )} & \ghost{U_S} & \ctrl{-3} & \qw & \qw\\
	 	 \lstick{ \ket 0^{\otimes 3} : } & \qw & \ghost{U_P \left ( \frac{\mu(x_{k}) - \mu^0}{\sqrt{1- \left (\sqrt{N}\bar \gamma \right )^2}} \right )} & \qw & \ghost{U_P \left (\frac{1}{\mu^0} \right )} & \ghost{U_S} & \qw & \qw & \qw\\
		\\ }}
\caption{Quantum circuit for solving the incremental homogenisation problem discretised with a uniform grid of~$N=2^3$ cells. The left most gate~$U_I$ encodes the prescribed average strain~$\bar \gamma$ as one of the two amplitudes of the~$\ket b$ qubit. The two gates~$U_P$ approximately implement the functions in their arguments. The incremental problem is solved by scaling the Fourier coefficients using the right~$U_P$ gate. The gate~$U_S$ imposes in the Fourier space the average prescribed strain. }
\label{fig:QRVE_inc_circ}
\end{figure}
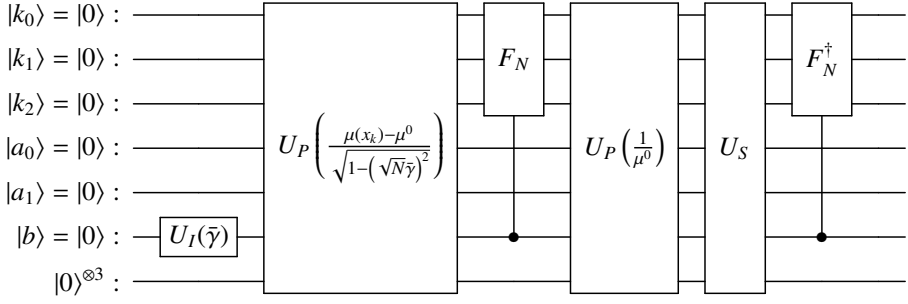
We are now ready to introduce the quantum circuit for solving the incremental problem in iteration step~$s$, see Algorithm~\ref{alg:DFT} and circuit diagram in Figure~\ref{fig:QRVE_inc_circ}. Step (a) of the algorithm is accomplished using the mapping
\begin{equation} \label{eq:upPolStress}
	U_P \left ( \frac{\mu(x_{k}) - \mu^0}{\sqrt{1- \left (\sqrt{N}\bar \gamma \right )^2}} \right ) \colon \ket k \ket 0 \ket 0^{\otimes (n+2)} \mapsto \sqrt{1-\frac{ \left (\mu(x_j)-\mu^0 \right )^2}{1- \left (\sqrt{N}\bar{\gamma} \right )^2}} \ket k \ket 0 \ket 0^{\otimes (n+2)} +\frac{\mu(x_j)-\mu^0}{\sqrt{1-\left (\sqrt{N}\bar{\gamma} \right )^2}} \ket k \ket 1 \ket 0^{\otimes (n+2)} \, .
\end{equation}
As introduced in Section~\ref{sec:encode_function}, the unitary~$U_P$ encodes the function in its argument by first approximating it as a piecewise polynomial and then encoding them as a sequence of controlled rotations. The unitary~$U_P$ is applied to the field qubits of~$\ket{q}^{(0)}_2$ when the AS qubit is in state~$\ket{b=0}$. The resulting state reads 
\begin{equation} \label{eq:psi3}
	\ket{q}^{(s)}_3 =\sqrt{N}\bar{\gamma} \underbrace{\ket{0}^{\otimes n}}_\text{field}
	\underbrace{\ket {0} \ket{0}}_\text{controls} 
	\underbrace{\ket {1}}_\text{AS}
	\underbrace{\ket 0^{\otimes n}}_\text{ancillas}
	+ \underbrace{ \sum_{k=0}^{2^n-1} \tau_k^{(s)} \ket k}_\text{field}
	\underbrace{\ket {1} \ket{0}}_\text{controls} 
	\underbrace{\ket {0}}_\text{AS}
	\underbrace{\ket 0^{\otimes n}}_\text{ancillas} + JUNK \, .
\end{equation} 
Here, $JUNK$ denotes the terms not further needed. The subsequent application of QFT, by applying the unitary~$F_N$ to the field qubits, gives the Fourier space representation~$\widehat{\ket{\tau}}^{(s)}$ of the polarisation stress, i.~e., 
\begin{equation} \label{eq:psi4}
	\ket{q}^{(s)}_4 =\sqrt{N}\bar{\gamma} \underbrace{\ket{0}^{\otimes n}}_\text{field}
	\underbrace{\ket {0} \ket{0}}_\text{controls} 
	\underbrace{\ket {1}}_\text{AS}
	\underbrace{\ket 0^{\otimes n}}_\text{ancillas}
	+ \underbrace{ \sum_{k=0}^{2^n-1} \hat \tau_k^{(s)} \ket k}_\text{field}
	\underbrace{\ket {1} \ket{0}}_\text{controls} 
	\underbrace{\ket {0}}_\text{AS}
	\underbrace{\ket 0^{\otimes n}}_\text{ancillas} + JUNK \, .
\end{equation} 
We can now compute the strain $\hat{\gamma}^{(s+1)}$. Note that for a one-dimensional problem the strain in the Fourier space is given by 
\begin{equation} 
\ket{\hat{\gamma}} = -\frac{1}{\mu^0} \ket{\hat{\tau}} \, ; 
\end{equation}
cf.~\eqref{eq:RVEsolution-2}. This operation is implemented using one more time the unitary mapping 
\begin{equation}
	U_P \left ( -\frac{1}{\mu^0} \right ) \colon \ket k \ket 1 \ket 0 \ket 0^{\otimes (n+1)}  \mapsto \sqrt{1-\frac{1}{\left (\mu^0 \right )^2}}\ket k \ket 1 \ket 0 \ket 0^{\otimes (n+1)}  -\frac{1}{\mu^0} \ket k \ket 1 \ket 1 \ket 0^{\otimes (n+1)}  \, .
\end{equation}
We note, although we have used the same symbol~$U_P$ above and in~\eqref{eq:upPolStress}, the two unitaries approximate different functions and their implementation details are different. The unitary~$U_P$ is applied to the field qubits when the first controls qubit is in state~$\ket{a_0 = 1}$, yielding the state
\begin{equation} \label{eq:psi4}
	\ket{q}^{(s)}_5 =\sqrt{N}\bar{\gamma} \underbrace{\ket{0}^{\otimes n}}_\text{field}
	\underbrace{\ket {0} \ket{0}}_\text{controls} 
	\underbrace{\ket {1}}_\text{AS}
	\underbrace{\ket 0^{\otimes n}}_\text{ancillas}
	- \underbrace{ \sum_{k=0}^{2^n-1} \frac{\hat \tau_k^{(s)}}{\mu^0} \ket k}_\text{field}
	\underbrace{\ket {1} \ket{1}}_\text{controls} 
	\underbrace{\ket {0}}_\text{AS}
	\underbrace{\ket 0^{\otimes n}}_\text{ancillas} + JUNK \, .
\end{equation} 
To impose the average strain~$\overline \gamma$, we use the unitary~$U_S$ introduced in Section~\ref{sec: base swap} to swap two of the amplitudes, with the result
\begin{multline}
	U_S \colon \sqrt{N}\bar{\gamma}\ket{0}^{\otimes n} \ket {0} \ket{0} \ket {1} \ket {0}^{\otimes n} - \frac{\hat \tau_0^{(s)}}{\mu^0}  \ket{k=0} \ket {1} \ket{0} \ket {0} \ket {0}^{\otimes n} + \dotsc \\ 
	\mapsto - \frac{\hat \tau_0^{(s)}}{\mu^0}  \ket{0}^{\otimes n} \ket {0} \ket{0} \ket {1} \ket {0}^{\otimes n} + \sqrt{N}\bar{\gamma} \ket{k=0} \ket {1} \ket{0} \ket {0} \ket {0}^{\otimes n} + \dotsc \, .
\end{multline}
After the amplitude swap, the quantum state becomes 
\begin{equation} \label{eq:psi4}
	\ket{q}^{(s)}_6 = - \frac{\hat \tau_0^{(s)}}{\mu^0} \underbrace{\ket{0}^{\otimes n}}_\text{field}
	\underbrace{\ket {0} \ket{0}}_\text{controls} 
	\underbrace{\ket {1}}_\text{AS}
	\underbrace{\ket 0^{\otimes n}}_\text{ancillas}
	+
	\underbrace{ \overline \gamma \ket k}_\text{field}
	\underbrace{\ket {1} \ket{1}}_\text{controls} 
	\underbrace{\ket {0}}_\text{AS}
	\underbrace{\ket 0^{\otimes n}}_\text{ancillas}
	- \underbrace{ \sum_{k=1}^{2^n-1} \frac{\hat \tau_k^{(s)}}{\mu^0} \ket k}_\text{field}
	\underbrace{\ket {1} \ket{1}}_\text{controls} 
	\underbrace{\ket {0}}_\text{AS}
	\underbrace{\ket 0^{\otimes n}}_\text{ancillas} + JUNK \, .
\end{equation} 
Finally, we compute the inverse QFT by applying~$F_N^\dagger$ to the field qubits, giving the desired state 
\begin{equation} \label{eq:psi4}
	\ket{q}^{(s)}_7 = \underbrace{ \sum_{k=0}^{2^n-1} \gamma_k^{(s+1)} \ket k}_\text{field}
	\underbrace{\ket {1} \ket{1}}_\text{controls} 
	\underbrace{\ket {0}}_\text{AS}
	\underbrace{\ket 0^{\otimes n}}_\text{ancillas} + JUNK \, .
\end{equation} 
The solution~$\ket {\gamma}^{(s+1)}$ and its components can then be obtained by measuring, as outlined in Section~\ref{sec:measurement}. When measuring the controls and AS qubits can be found either in state $\ket 0$ or~$\ket 1$. The ancillas are all in state~$\ket 0^{\otimes n}$, cf. Sections~\ref{sec:multi-controll} and~\ref{sec:encode_function}.
%
\subsubsection{Fixed-point iteration}
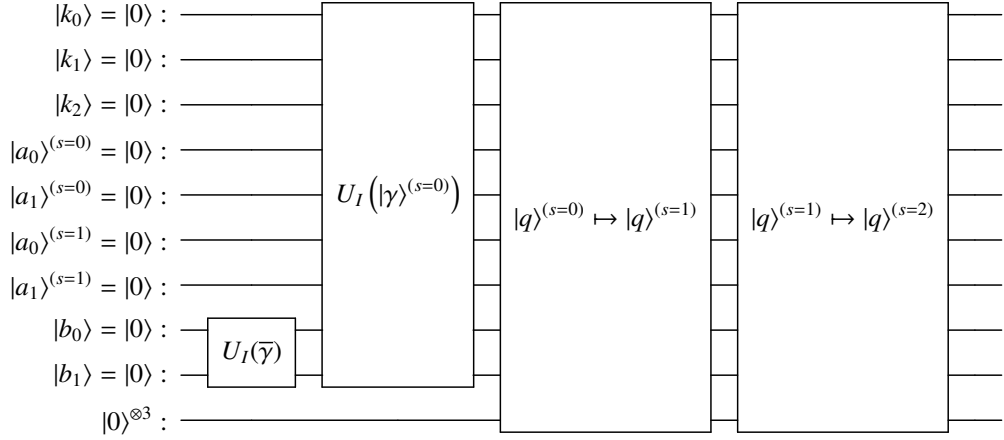
\begin{figure}
\centering
\scalebox{1.0}{
\Qcircuit @C=1.0em @R=0.8em @!R { \\
	 	 \lstick{\ket{k_0} = \ket 0 : } & \qw & \multigate{8}{ U_I \left( \ket{\gamma}^{(s=0)} \right ) } & \multigate{9}{\ket q^{(s=0)} \mapsto \ket q^{(s=1)} } & \multigate{9}{\ket q^{(s=1)} \mapsto \ket q^{(s=2)}} & \qw & \qw\\
	 	 \lstick{\ket{k_1} = \ket 0 : } & \qw & \ghost{U_I \left( \ket{\gamma}^{(s=0)} \right )} & \ghost{\ket q^{(s=0)} \mapsto \ket q^{(s=1)}} & \ghost{\ket q^{(s=1)} \mapsto \ket q^{(s=2)}} & \qw & \qw\\
	 	 \lstick{\ket{k_2} = \ket 0 :} & \qw & \ghost{\mathrm{Initialize\,\gamma_0}} & \ghost{\ket q^{(s=0)} \mapsto \ket q^{(s=1)}} & \ghost{\ket q^{(s=1)} \mapsto \ket q^{(s=2)}} & \qw & \qw\\
	 	 \lstick{\ket{a_0}^{(s=0)} = \ket 0 : } & \qw & \ghost{U_I \left( \ket{\gamma}^{(s=0)} \right ) } & \ghost{\ket q^{(s=0)} \mapsto \ket q^{(s=1)}} & \ghost{\ket q^{(s=1)} \mapsto \ket q^{(s=2)}}  & \qw & \qw\\
	 	 \lstick{\ket{a_1}^{(s=0)} = \ket 0 : } & \qw & \ghost{U_I \left( \ket{\gamma}^{(s=0)} \right ) } & \ghost{\ket q^{(s=0)} \mapsto \ket q^{(s=1)}} &  \ghost{\ket q^{(s=1)} \mapsto \ket q^{(s=2)}}  & \qw & \qw\\
	 	 \lstick{\ket{a_0}^{(s=1)} = \ket 0: } & \qw & \ghost{U_I \left( \ket{\gamma}^{(s=0)} \right ) } & \ghost{\ket q^{(s=0)} \mapsto \ket q^{(s=1)}} & \ghost{\ket q^{(s=1)} \mapsto \ket q^{(s=2)}}  & \qw & \qw\\
	 	 \lstick{\ket{a_1}^{(s=1)} = \ket 0: } & \qw & \ghost{U_I \left( \ket{\gamma}^{(s=0)} \right )} & \ghost{\ket q^{(s=0)} \mapsto \ket q^{(s=1)}} &  \ghost{\ket q^{(s=1)} \mapsto \ket q^{(s=2)}}  & \qw & \qw\\
	 	 \lstick{\ket{{b}_{0}}  = \ket 0: } & \multigate{1}{U_I(\overline \gamma) } & \ghost{\mathrm{Initialize\,\gamma_0}} & \ghost{\ket q^{(s=0)} \mapsto \ket q^{(s=1)}} &  \ghost{\ket q^{(s=1)} \mapsto \ket q^{(s=2)}} & \qw & \qw\\
	 	 \lstick{\ket{{b}_{1}} = \ket 0: } &  \ghost{U_I(\overline \gamma) }  & \ghost{\mathrm{Initialize\,\gamma_0}} & \ghost{\ket q^{(s=0)} \mapsto \ket q^{(s=1)}}  & \ghost{\ket q^{(s=1)} \mapsto \ket q^{(s=2)}}  & \qw & \qw\\
	 	 \lstick{\ket {0}^{\otimes 3}: } & \qw & \qw & \ghost{\ket q^{(s=0)} \mapsto \ket q^{(s=1)}} & \ghost{\ket q^{(s=1)} \mapsto \ket q^{(s=2)}}  & \qw & \qw\\
\\ }}
\caption{Schematic of the quantum circuit for iterative solution of the periodic homogenisation problem discretised with a uniform grid of~$N=2^3$  cells using two iteration steps~($S=2$) . The two right most gates implementing~$\ket{q}^{(s)} \mapsto \ket{q}^{(s+1)} $ follow the the quantum circuit for solving the incremental problem shown in Figure~\ref{fig:QRVE_inc_circ}.}
\label{fig:QRVE_all}
\end{figure}

%
Finally, we turn to the implementation of the fixed-point iteration according to Algorithm \ref{alg:DFT}. A schematic of the proposed circuit is shown in Figure~\ref{fig:QRVE_all}. We expand the quantum state vector~\eqref{eq:statOneIterativeStep} as follows  
\begin{equation}
	\ket q = \underbrace{\ket {0}^{\otimes n} }_\text{field}
\underbrace{ \ket{a_0} \ket{a_1} \ket{a_2} \ket{a_3 } \cdots \ket{a_{2S-2}} \ket{a_{2S-1}}}_\text{controls} 
\underbrace{\ket {b_0} \ket {b_1} \dotsc \ket {b_{S-1}} }_\text{AS}
\underbrace{{\ket 0}^{\otimes n}}_\text{ancillas}  \, ,  
\end{equation}
where~$S$ is the total number of iterations and the definition of the qubits is analogous to the previous section. The number of the qubits for controls and encoding the applied strain (AS) depend on the number of iterations.    The first pair of the control qubits~$\ket {a_0} \ket {a_1}$ is  for the first iteration step, the second pair~$\ket {a_2} \ket {a_3}$ for the second iteration step and so on. Furthermore, we store $S$copies of the applied strain~$\overline \gamma \in \mathbb R$ because of the no-cloning theorem of  quantum mechanics. 

As depicted in Figure~\ref{fig:QRVE_all}, we begin by  encoding the applied strain~$\overline \gamma$ so that the initial quantum state becomes
\begin{equation} 
\ket{q}^{(s=0)} =   \sqrt{1-S \overline{\gamma}^2}  \underbrace{\ket {0}^{\otimes n} }_\text{field}
\underbrace{ \ket{0}^{\otimes 2S}}_\text{controls} 
\underbrace{\ket {0}^{\otimes S}}_\text{AS}
\underbrace{{\ket 0}^{\otimes n}}_\text{ancillas}  + \sum_{s=1}^{S-1} \overline \gamma \underbrace{\ket {0}^{\otimes n} }_\text{field}
\underbrace{ \ket{0}^{\otimes 2S}}_\text{controls} 
\underbrace{\ket {s}}_\text{AS}
\underbrace{{\ket 0}^{\otimes n}}_\text{ancillas}  \, .
\end{equation} 
Subsequently, the field qubits are initialised with the predictor strain $\ket{\gamma}^{(s=0) } \in \mathbb R^{N}$. We choose~$\ket{\gamma}^{(s=0) } = \frac{1}{\sqrt{N}} \sum \ket k $. To implement the fixed-point iteration, we apply repeatedly the previously introduced quantum algorithm for solving the incremental problem, see Figure~\ref{fig:QRVE_inc_circ}. The state of the quantum system after the first iteration is given by
\begin{equation} 
\ket{q}^{(s=1)} =    \underbrace{   \sum_{k=0}^{2^n-1} \gamma_k^{(s=1)} \ket {k} }_\text{field}
\underbrace{ \ket{1} \ket{1} \ket{0} \ket{0 } \cdots \ket{0} \ket{0}}_\text{controls} 
\underbrace{\ket {0}^{\otimes S}}_\text{AS}
\underbrace{{\ket 0}^{\otimes n}}_\text{ancillas}  + \sum_{s=0}^{S-1} \overline \gamma \underbrace{\ket {0}^{\otimes n} }_\text{field}
\underbrace{ \ket{0}^{\otimes 2S}}_\text{controls} 
\underbrace{\ket {s}}_\text{AS}
\underbrace{{\ket 0}^{\otimes n}}_\text{ancillas} + JUNK \, .
\end{equation} 
In the second step, the quantum algorithm for solving the incremental problem is applied only when the first pair of controls qubits are in the state~$\ket{a_0} = \ket{a_1} = \ket 1$, yielding the state vector
\begin{equation} 
\ket{q}^{(s=2)} =    \underbrace{   \sum_{k=0}^{2^n-1} \gamma_k^{(s=2)} \ket {k} }_\text{field}
\underbrace{ \ket{1} \ket{1} \ket{1} \ket{1 } \cdots \ket{0} \ket{0}}_\text{controls} 
\underbrace{\ket {0}^{\otimes S}}_\text{AS}
\underbrace{{\ket 0}^{\otimes n}}_\text{ancillas}  + \sum_{s=0}^{S-1} \overline \gamma \underbrace{\ket {0}^{\otimes n} }_\text{field}
\underbrace{ \ket{0}^{\otimes 2S}}_\text{controls} 
\underbrace{\ket {s}}_\text{AS}
\underbrace{{\ket 0}^{\otimes n}}_\text{ancillas} + JUNK \, .
\end{equation} 
Hence, after~$S$ iteration steps, the final quantum state vector is given by 
\begin{equation} 
\label{eq:final state}
\ket{q}^{(s=S)} =    \underbrace{   \sum_{k=0}^{2^n-1} \gamma_k^{(s=S)} \ket {k} }_\text{field}
\underbrace{ \ket{1} \ket{1} \ket{1} \ket{1 } \cdots \ket{1} \ket{1}}_\text{controls} 
\underbrace{\ket {0}^{\otimes S}}_\text{AS}
\underbrace{{\ket 0}^{\otimes n}}_\text{ancillas}  +  JUNK \, .
\end{equation} 
Notice that all controls qubits are in the state~$\ket 1$. It bears emphasis that the incremental circuit for iteration $s$ will be applied if and only if the control qubits from iteration $s-1$ are all in the state $\ket 1$. The algorithm is successful if all control qubits are in the state \(|1\rangle\) and all AS qubits are in the state \(|0\rangle\). Consequently, the total number of~\mbox{$JUNK$} terms is given by \((3S^2+S-1)2^n\).

\begin{figure} 
\centering
\includegraphics[width=0.8\textwidth]{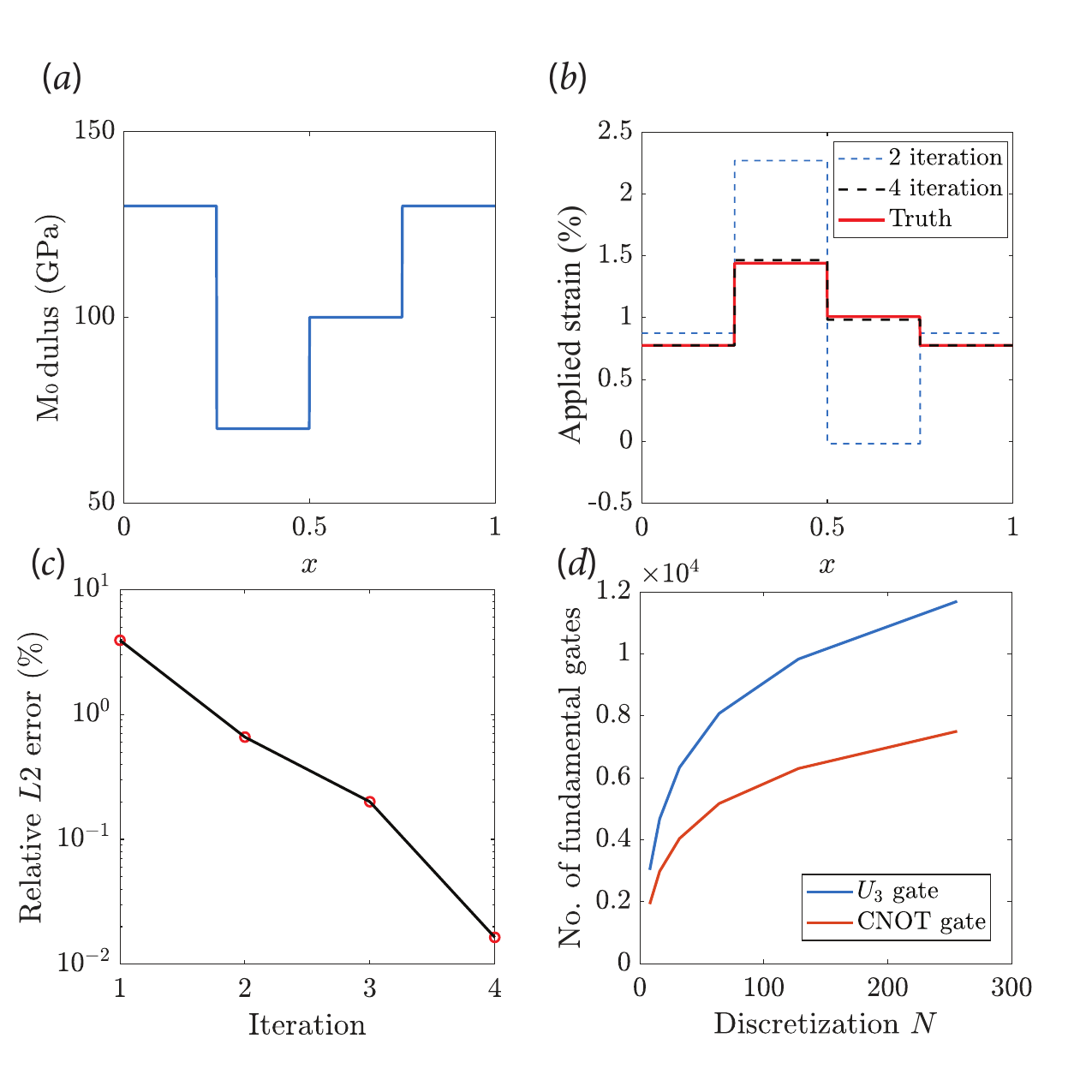}
\caption{Periodic homogenisation of a  one-dimensional RVE with a domain~$\Omega=(0, \,1)$. (a) Piecewise constant shear modulus~$\mu (x)$. (b) True strain field~$\gamma(x)$ and computed strain fields at steps $s=2$ and~$s=4$.  (d) Convergence of the relative $L_2$-norm error. (d) Scaling of the total number of $U_3$  and $CNOT$ gates in the quantum circuit. }
\label{fig:QRVE}
\end{figure}

The solution of a one-dimensional homogenisation problem using the circuit just described is shown in Figure~\ref{fig:QRVE}. The convergence of the computed strain field in Figure~\ref{fig:QRVE}(b) is indicative of rapid convergence towards the exact solution. The rate of convergence is quantified in Figure~\ref{fig:QRVE}(c), which is suggestive of exponential convergence.

\paragraph{Complexity} Finally, we assess the complexity of the algorithm by counting the number of~$U_3$ and $CNOT$ gates in the circuit as a function of the number of grid cells~$N$, see Figure~\ref{fig:QRVE}(d). It is evident that the scaling is polylogarithmic, meaning that only $\poly \log (N)$ universal gates are required, hence proving the efficiency of the algorithm and circuit design. 

%

%
\section{Summary and concluding remarks \label{sec:conclusions}}
%

We have carried out exploratory work aimed at identifying opportunities for the application of quantum computing concepts and techniques to solid mechanics. Specifically, on the basis of simple test cases our work suggests that quantum computing can indeed accelerate exponentially typical representative volume element (RVE) calculations in computational homogenization, thus potentially bringing concurrent multiscale computing within reach of practicality. 

However, the desired exponential speedup is achieved at the expense of a complete and fundamentally new redesign and overhaul of the classical algorithms and solvers. Thus, from a representational standpoint the physical variables of the problem must be encoded as the amplitudes of an entangled quantum system, a complete paradigmatic departure from the binary arithmetic and bit operations of classical digital computers. We have presented two state preparation algorithms, one approximate and one exact, that can be used to encode RVE information as a quantum state and which demonstrate the feasibility of the representation. The remarkable benefit of this representational paradigm shift is that the space of entangled quantum states is exponentially larger than the classical configuration space spanned by the same number of particles, or qubits, which constitutes a first remarkable game-changer. 

Since quantum computers are governed by the laws of quantum mechanics, the state of the system must be updated---and evolved towards the solution---through the application of unitary transformations, specifically elementary unitary operations or `gates'. We have illustrated the feasibility of this quantum implementation in the context of RVE solvers through the design of a number of algorithms and circuits, including the Quantum Fourier Transform, polynomial encoding and fixed-point iteration. The payoff of this reformulation is that the quantum computer, as is typical of analog computers, updates the entire state of the system in one fell swoop. The resulting quantum parallelism and interference provides the second and definitive game-changer. 

We close by emphasising that, owing to current limitations in the available quantum platforms, including emulators, tests are of necessity limited in size and complexity. We expect that full-fledged two and three-dimensional, linear and non-linear, static and dynamic implementations of RVE solvers and other similar applications will become possible, including noise and error control, as the quantum platforms improve in power and accessibility. Also to be expected is the development of basic utility libraries facilitating the programming of quantum computers and eschewing the need for circuit design at the gate level. In view of the tremendous promise of the field, these and other developments suggest themselves as worthwhile undertakings at the intersection between quantum computing and computational mechanics. 

\section*{Acknowledgements}

M.~Ortiz gratefully acknowledges the support of the Deutsche Forschungsgemeinschaft (DFG, German Research Foundation) {\sl via} Germany's Excellence Strategy (GZ 2047/1,  project 390685813) and the Hausdorff Center for Mathematics of Bonn University. B.~Liu gratefully acknowledges the support of Granta Design faculty startup fund from the Engineering Department of the University of Cambridge.



\bibliographystyle{plain}
\bibliography{qFEM}

\end{document}